\documentclass[lettersize,journal]{IEEEtran}
\usepackage{amsmath,amsfonts}
\usepackage{algorithmic}
\usepackage{algorithm}
\usepackage{array}
\usepackage[caption=false,font=normalsize,labelfont=sf,textfont=sf]{subfig}
\usepackage{textcomp}
\usepackage{stfloats}
\usepackage{url}
\usepackage{verbatim}
\usepackage{graphicx}
\usepackage{cite}
\usepackage{lineno}
\usepackage{multirow}
\usepackage{float}
\usepackage{graphicx}
\usepackage{amsmath}
\usepackage{color}
\usepackage{soul}
\usepackage{booktabs}
\usepackage{threeparttable}
\usepackage{caption}
\usepackage{indentfirst}
\usepackage[T1]{fontenc}

\begin{document}

\title{Sense: Model-Hardware Co-design for Accelerating Sparse CNNs on Systolic Array}

\author{Wenhao~Sun,
	Deng~Liu,
	Zhiwei~Zou,
	Wendi~Sun,
	Song~Chen,~\IEEEmembership{Member,~IEEE,}
	and~Yi~Kang

}

\maketitle
\begin{abstract}
	\textbf{Sparsity is an intrinsic property of convolutional neural network(CNN) and worth exploiting for CNN accelerators, but extra processing comes with hardware overhead, causing many architectures suffering from only minor profit. Meanwhile, systolic array has been increasingly competitive on CNNs acceleration for its high spatiotemporal locality and low hardware overhead. However, the irregularity of sparsity induces imbalanced workload under the rigid systolic dataflow, causing performance degradation. Thus, this paper proposed a systolic-array-based architecture, called Sense, for sparse CNN acceleration by model-hardware co-design,  \iffalse for both sparse input feature map(IFM) and weight processing,\fi achieving large performance improvement. To balance input feature map(IFM) and weight loads across Processing Element(PE) array, we applied channel clustering to gather IFMs with approximate sparsity for array computation, and co-designed a load-balancing weight pruning method to keep the sparsity ratio of each kernel at a certain value with little accuracy loss, improving PE utilization and overall performance. Additionally, Adaptive Dataflow Configuration is applied to determine the computing strategy based on the storage ratio of IFMs and weights, lowering \boldmath{$1.17\times$\textasciitilde$1.8\times$} DRAM access compared with Swallow and further reducing system energy consumption. The whole design is implemented on ZynqZCU102 with 200MHz and performs at 471-, 34-, 53- and 191-image/s for AlexNet, VGG-16, ResNet-50 and GoogleNet respectively. Compared against sparse systolic-array-based accelerators, Swallow, FESA and SPOTS,  Sense achieves \boldmath{$1\times$\textasciitilde$2.25\times$}, \boldmath{$1.95\times$\textasciitilde$2.5\times$} and \boldmath{$1.17\times$\textasciitilde$2.37\times$} performance improvement on these CNNs respectively with reasonable overhead.}
\end{abstract}


\section{Introduction}

Nowadays, neural network has been widely applied in numerous domains, such as image recognition\cite{10.1145/3065386}, speech recognition\cite{7178838}, object detection\cite{7780460}, computer vision\cite{10.1145/2647868.2654889}, etc. Especially, CNNs \cite{10.1145/3065386},\cite{simonyan2015deep},\cite{7298594},\cite{7780459},\cite{7780460} have outstood in image recognition and object detection. As higher precision and complication demands come up, the workloads of network computing are still growing. Therefore, researchers have been bending their efforts for exploration of neural network hardware accelerators to catch up with software development.	\\ 
\indent\setlength{\parindent}{1em}Although CNNs bring intensive computations, there exists monotonous operations; thus, many hardware architects attempt to reduce runtime by improving computation parallelism. Meanwhile, weights and IFMs can be massively reused across operations. Since data movement consumes much more energy than computation\cite{10.1145/1816038.1815968}, especially with DRAM, many sophisticated dataflow strategies were designed to improve data reuse rate. For example, Eyeriss\cite{7738524} applies row stationary dataflow to minimize movement consumption; TPU\cite{8192463} revives systolic array to realize high energy efficiency and throughput with structured dataflow and low bandwidth; Thinker\cite{8207783} supports bit-width adaptive computing and on-demand array partitioning and reconfiguration, to maximize resource utilization. These works mainly actualize acceleration through parallelism improvement and dataflow optimization, without consideration of sparsity processing on IFMs and weights, causing redundant storage and computation. \\ 
\indent\setlength{\parindent}{1em}Furthermore, sparsity is an intrinsic property of CNNs. For IFMs, approximately half of their elements are zeros because of the widely used nonlinear function ReLU; for weights, the fault tolerance of CNN allows of a large number of redundant zeros, which inspires researchers to develop pruning and quantization methods\cite{park2017faster},\cite{8574528}, \cite{6639346},\cite{10.5555/3295222.3295239},\cite{li2017pruning},\cite{zhu2017prune},\cite{frankle2019lottery}. Therefore, various sparsity-aware CNN accelerators were designed. EIE\cite{7551397} supports both sparse weight and IFM processing, but only focused on fully-connected(FC) layers. Cambricon-X\cite{7783723} and Cnvlutin\cite{7551378}, respectively, take advantage of sparse weights and IFMs to eliminate operations induced by zero. SCNN\cite{8192478} can support arbitrary sparse patterns, but it fails to fully leverage the beneﬁts provided by sparse IFMs and weights due to imbalanced workload and access contention across independent PEs. Lu $et$ $al$\cite{9130762} proposed an FPGA architecture to deal with the irregular connection in sparse convolution(CONV) layers, but huge LUT resource was consumed to match the indexes between weights and IFMs and increases linearly with PE array scaling, which becomes the bottleneck of performance. Zhu $et$ $al$\cite{8735526} designed an FPGA accelerator to avoid extra coordinate computation for connection reconstructing or output locating with a sparsewise dataflow. However, since this accelerator consumes large BRAM resource for output feature map(OFM) storage in each PE, the performance is bounded by the number of BRAM on FPGA. Therefore, these architectures suffer from one-side sparsity exploration, access contention or inefficient resource utilization. \\ 
\indent\setlength{\parindent}{1em}Since systolic array\cite{1653825} is widely applied to accelerate CNNs\cite{8192463},\cite{8207783},\cite{7870350},\cite{6881431} for higher energy efficiency and reource efficiency compared with other architectures as shown in Tab.\ref{tab:tab1}. Thus, researchers try to process sparsity with systolic architecture to improve the overall benefits. Swallow\cite{9026967} overcomes the inability to exploit the sparsity of weights and IFMs, CONV layers and FC layers of CNNs with limited resource in a systolic array, and introduce a sparse-aware dataflow to boost PE utilization, achieving higher bandwidth saving and energy efficiency compared with previous sparse accelerators. However, the structured systolic dataflow essentially contradicts with the irregularity of sparsity, causing imbalanced PE loads. Considering that, FESA\cite{9218630} pruned the kernels to $2$\textasciitilde$7$ formalized zero distribution patterns and left IFM unprocessed to regularize dataflow as the dense systolic tempo, achieving lower sparsity processing overhead. But this pruning method is only implemented on Cifar-10 and Cifar-100\cite{article} currently. Thus, to balance workload with higher versatility in systolic array, SPOTS\cite{10.1145/3532863} designed a group-wise pruning method to divide weights into groups and prune some elements of the same position in each group, which achieves similar versatility with shape-wise pruning method\cite{10.5555/3157096.3157329} and improves compatibility with systolic array. Accordingly, SPOTS applied Image to Column (Im2Col) transformation of IFMs coupled with general matrix-matrix multiplication (GEMM) to better fit its pruning scheme into systolic array by skipping the weight rows and IFM columns with all zeros. However, since its pruning method is too fine-grained, the sparsity of weights after pruning is bounded by accuracy. Besides, SPOTS fails to exploit the sparsity in those rows and columns with some zeros, causing inefficient acceleration.

\begin{table}[]
	\centering
	\caption{Comparison of systolic accelerator against others}
	\label{tab:tab1}
	\begin{tabular}{|c|c|c|c|}
		\hline
		Accelerator            & \begin{tabular}[c]{@{}c@{}}Technology\\ (nm)\end{tabular} & \begin{tabular}[c]{@{}c@{}}Resource\\ Efficiency\\ (GOPs/mm2)\end{tabular} & \begin{tabular}[c]{@{}c@{}}Energy\\ Efficiency\\ (GOPs/W)\end{tabular} \\ \hline
		NVIDIA K80             & 28                                                        & 4.99                                                                       & 0.03@FP32                                                              \\ \hline
		DaDianNao\cite{7011421}              & 28                                                        & 83.3                                                                       & 0.34@INT16                                                             \\ \hline
		Eyeriss\cite{7738524}                & 65                                                        & 6.86                                                                       & 0.14@INT16                                                             \\ \hline
		\textbf{TPU(Systolic)\cite{8192463}} & 28                                                        & \textbf{278}                                                               & \textbf{2.3@IN8}                                                       \\ \hline
	\end{tabular}
\end{table}

\indent\setlength{\parindent}{1em}Additionally, memory access occupies a huge proportion of system energy consumption, making it critical to further reduce memory access through dataflow. But sparse IFMs and weights can be irregular and fragmented, which leads to lower memory access efficiency. SCNN\cite{8192478} employed a novel dataflow to eliminate unnecessary data transfers, but access contentions occurred when routing the products to accumulator buffer due to irregular sparse patterns. Lu $et$ $al$\cite{8735526} proposed a weight layout to enable efficient memory access without conflicts, but huge LUT consumption blocked the performance. Swallow harnesses a sparsity-aware dataflow with matrix multiplication tiling to promote data reuse within each channel, reducing DRAM access with little overhead. However, Swallow always preferentially reuse IFMs, while DRAM access can be variable if we choose different reuse strategies. Thus, there is still room to further lower DRAM access by choosing dataflow according to the storage ratio of IFMs and weights in each layer.  \\ 
\indent\setlength{\parindent}{1em}These previous sparse systolic accelerators suffered from imbalanced workload, lacking versatility or low sparsity of weight pruning.  Besides, the dataflow is inflexible for the variable ratio of IFM and weight in each layer. Thus, this paper aims to balance workload to fit with sparse systolic array, while maintaining the sparsity and versatility of weight pruning with reasonable overhead, and further optimize DRAM access. A model-hardware co-design of sparse CNN accelerator based on systolic array is proposed to improve system performance and energy efficiency. Our main contributions are as follows: \\ 
\indent\setlength{\parindent}{1em}1). We proposed a systolic-array-based accelerator for both sparse IFM and weight processing with reasonable overhead. It supports the whole kernel and IFM block compression and eliminates zero element operations to accelerate computing.\\ 
\indent\setlength{\parindent}{1em}2). We applied channel clustering to gather IFMs with approximate sparsity to compute in the PE array by sorting channel indexes according to the numbers of nonzero elements(NZEs) in IFMs, and co-designed a load-balancing weight pruning method to keep the sparsity ratio of each kernel at a certain value with little accuracy loss, which can balance PE loads across systolic array and greatly improve performance. \\ 
\indent\setlength{\parindent}{1em}3). We proposed an Adaptive Dataflow Configuration to optimize data reuse rate and further reduce DRAM access. The reuse strategy of our dataflow can be flexibly configured according to the storage ratio of IFMs and weights. \\ 
\indent\setlength{\parindent}{1em}Comparing with sparse systolic accelerator, Swallow\cite{9026967}, we balance the workload of sparse IFMs and weights in systolic array and achieve $1\times$\textasciitilde$2.25\times$ speedup, and $0.97\times$\textasciitilde$2.2\times$ energy efficiency with 1.6\% LUT, 7\%BRAM and 3\% power overhead. Based on the Adaptive Dataflow Configuration, we reduce $1.17\times$\textasciitilde$1.8\times$ DRAM access. Besides, comparing with the pruning method of FESA\cite{9218630}, we achieve $1.2\times$\textasciitilde$1.3\times$ speedup in terms of weights and implements on more complicated dataset, such as ImageNet, with higher versatility; by exploiting sparsity of both IFMs and weights, we achieve $1.95\times$\textasciitilde$2.5\times$ performance improvement with $1.5\times$ power. Further, compared with SPOTS\cite{10.1145/3532863}, our pruning method obtain $1\times$\textasciitilde$2.02\times$ sparsity of weights, achieving $1.17\times$\textasciitilde$2.37\times$ speedup with $1.3\times$ power. \\ 

\section{Preliminary}

According to the computing process of CNN, skipping zeros of IFMs and weights can effectively accelerate execution. However, for systolic array, there still exists problems of imbalanced workloads——that is, PEs with IFMs and weights of high sparsity rate(zero percentage) must wait for those with low sparsity rate to finish. Thus, lots of PE will be idle, reducing array utilization rate. Additionally, when a certain PE contains data with really low sparsity rate, it greatly block the overall acceleration. Consequently, the key challenge is to balance workloads across systolic array.

\subsection{Computing Process of CNN}

\begin{figure}[H]
	\centering
	\includegraphics[width=0.8\linewidth]{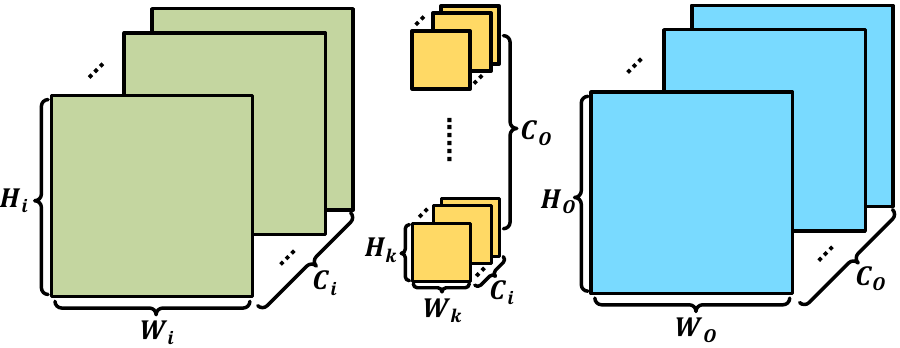}
	\caption{ Convolution process. }
	\label{fig1}
\end{figure}

The computing process of CNN is shown in Fig.\ref{fig1}. A CONV layer receives $C_i$ input channels(ICs) of IFMs with size of $H_i \times W_i$. Each OFM of $C_o$ output channels(OCs), with size of $H_o \times W_o$, is generated by sliding a $H_k \times W_k$ kernel window on one IFM and then accumulating across the temporary results of $C_i$ ICs. Finally, $C_o$ OFMs are calculated by convolving $C_o$ groups of kernels with shared IFMs. Set IFMs as matrix $I$[$C_i$,$H_{i}$,$W_{i}$], weights as $W$[$C_o$,$C_i$,$H_{k}$,$W_{k}$] and OFMs as $O$[$C_o$,$H_o$,$W_o$], the process can be described as Equ.\ref{equ3}.	
\begin{equation}
	\label{equ3}
	\begin{aligned}
		O[o,x,y] = \sum_{i=0}^{C_i} \sum_{a=0}^{H_k} \sum_{b=0}^{H_k} I[i,x+a,y+b] \times W[o,i,a,b] \\
		(0 \leq o < C_o, 0 \leq x < H_{o}, 0 \leq y < W_{o})
	\end{aligned}
\end{equation}

\subsection{Systolic Array}
\begin{figure}[H]
	\centering
	\includegraphics[width=0.65\linewidth]{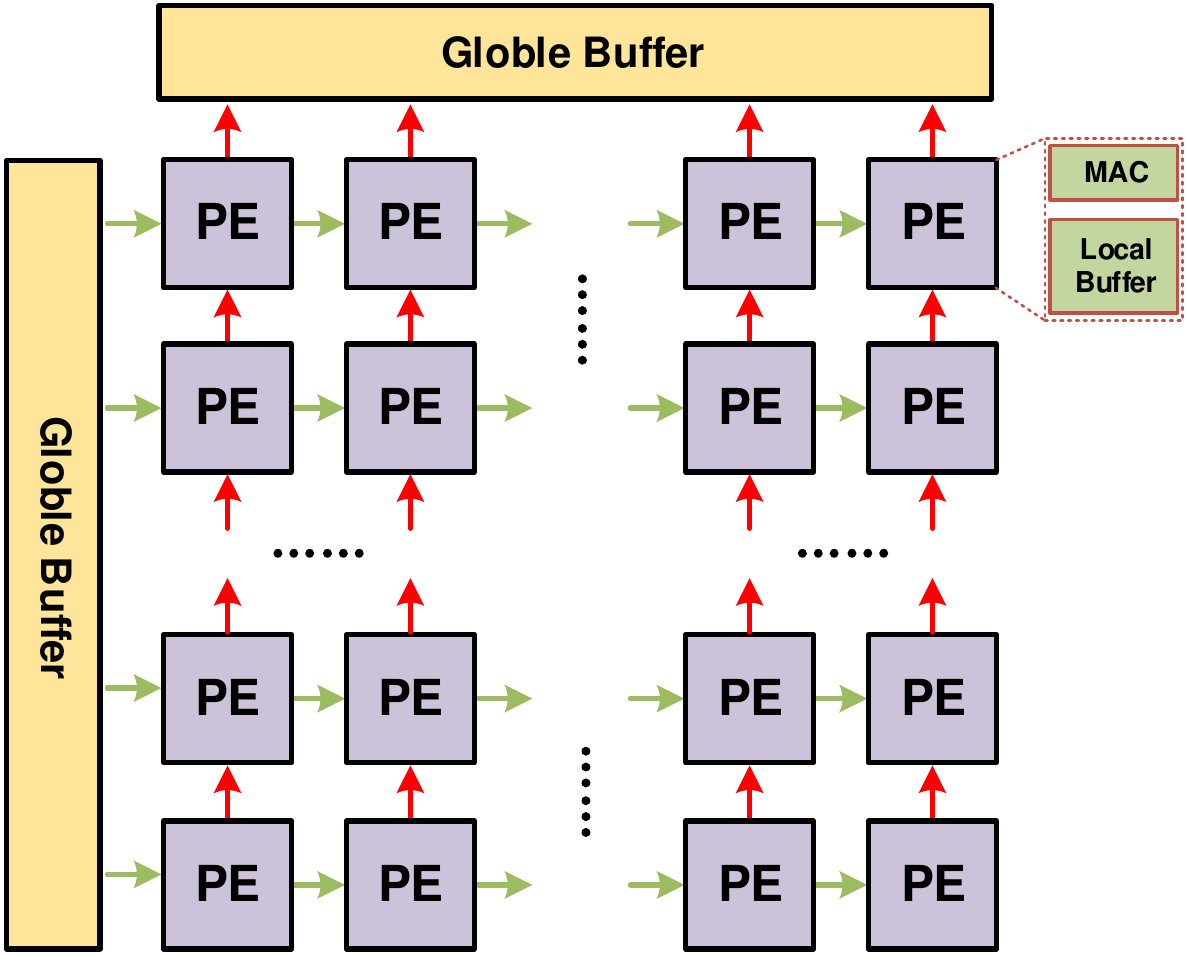}
	\caption{Systolic array architecture.}
	\label{fig2}
\end{figure}

Systolic array\cite{1653825} has been widely adopted for CNN acceleration\cite{8192463},\cite{8207783},\cite{7870350},\cite{6881431} with distinct advantages. Firstly, a systolic architecture consists of massive simple and uniform PEs, achieving high performance with low hardware cost. Secondly, systolic array propagates data horizontally and vertically with simplified inter-PE connection, which can highly reuse data and reduce date movement consumption. Finally, communication with the outside occurs only at the boundary PEs, providing huge bandwidth saving. An example of systolic array is shown in Fig.\ref{fig2}. Each PE conducts MAC operation with locality storage. Based on various dataflows, each PE can station one input of MAC temporally and transfer other inputs to neighbor PEs spatially. Due to the structured dataflow, control unit only takes up a small portion of system hardware cost, achieving high resource utilization. Despite the strong points mentioned above, there still exists challenges to accelerate sparse CNN on systolic array. 
\subsection{Challenges of Sparse Systolic Accelerator}
\begin{figure}[H]
	\centering
	\includegraphics[width=0.99\linewidth]{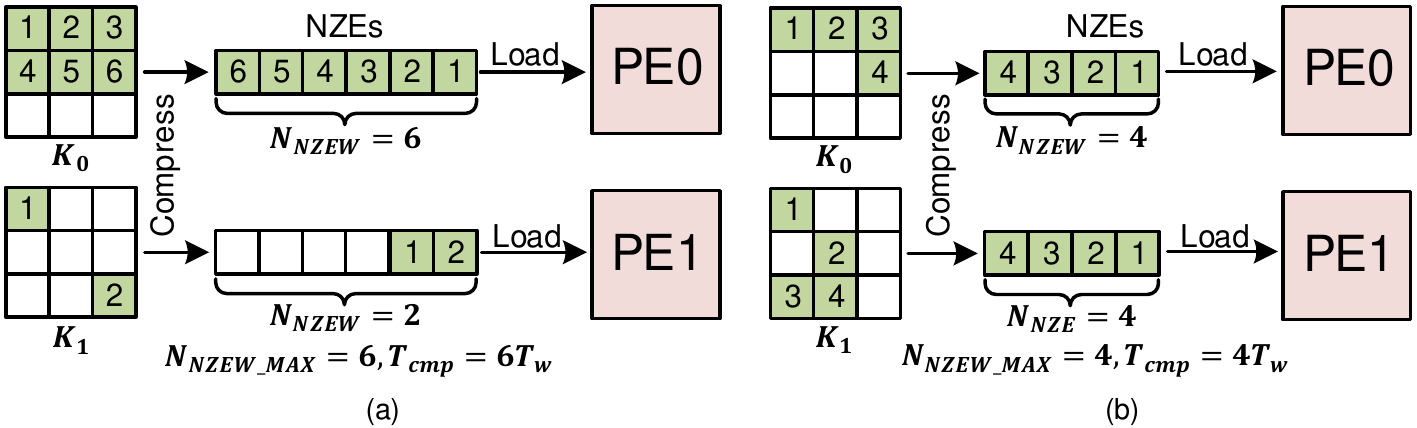}
	\caption{Imbalanced and balanced load of sparse weights in systolic array. (a)Imbalanced load.\iffalse The computing time of imbalanced workload is 6$T_w$ with 33\% PE idle rate.($T_w$: time for processing one weight element).\fi (b)Balanced load.\iffalse The computing time of imbalanced workload is 4$T_w$, achieving $1.5\times$x speedup.\fi}
	\label{fig3}
\end{figure}

To accelerate CNN processing, mapping computations on a systolic array can greatly improve parallelism; considering sparsity, when skipping zero computations, we manage to reduce computational complexity by cutting down the sizes of IFMs and kernels, further improving overall performance. However, the irregularity of sparsity conflict with the structured dataflow in systolic array, which may induce imbalanced workloads, causing low PE utilization and insufficient acceleration. Since CONV layers takes up the majority computation of CNNs, the challenges are analyzed from the perspective of weights and IFMs respectively in CONV layers.  \\
\indent\setlength{\parindent}{1em}For sparsity of weights, they contain a large number of redundant zeros after pruning, but the sparsity ratio of each kernel differs, causing imbalanced workloads across PE array. An example of computing flows towards weights with imbalanced loads after pruning is shown in Fig.\ref{fig3}(a). Kernel $K_0$ and $K_1$ are compressed and loaded in $PE_0$ and $PE_1$ respectively. Assume the number of NZEs in each kernel is $N_{NZEW}$ and the largest value is $N_{NZEW\_MAX}$, and the computing time needed for a single weight is $T_W$, the total computing time of $PE_0$ and $PE_1$ are $6T_W$ and $2T_W$ respectively. Based on the regular dataflow of systolic array, the computing time of the whole system is blocked to $6T_W$ and $PE_1$ will be idle for $4T_W$, indicating low performance and PE utilization rate.

\begin{figure}[H]
	\centering
	\includegraphics[width=0.99\linewidth]{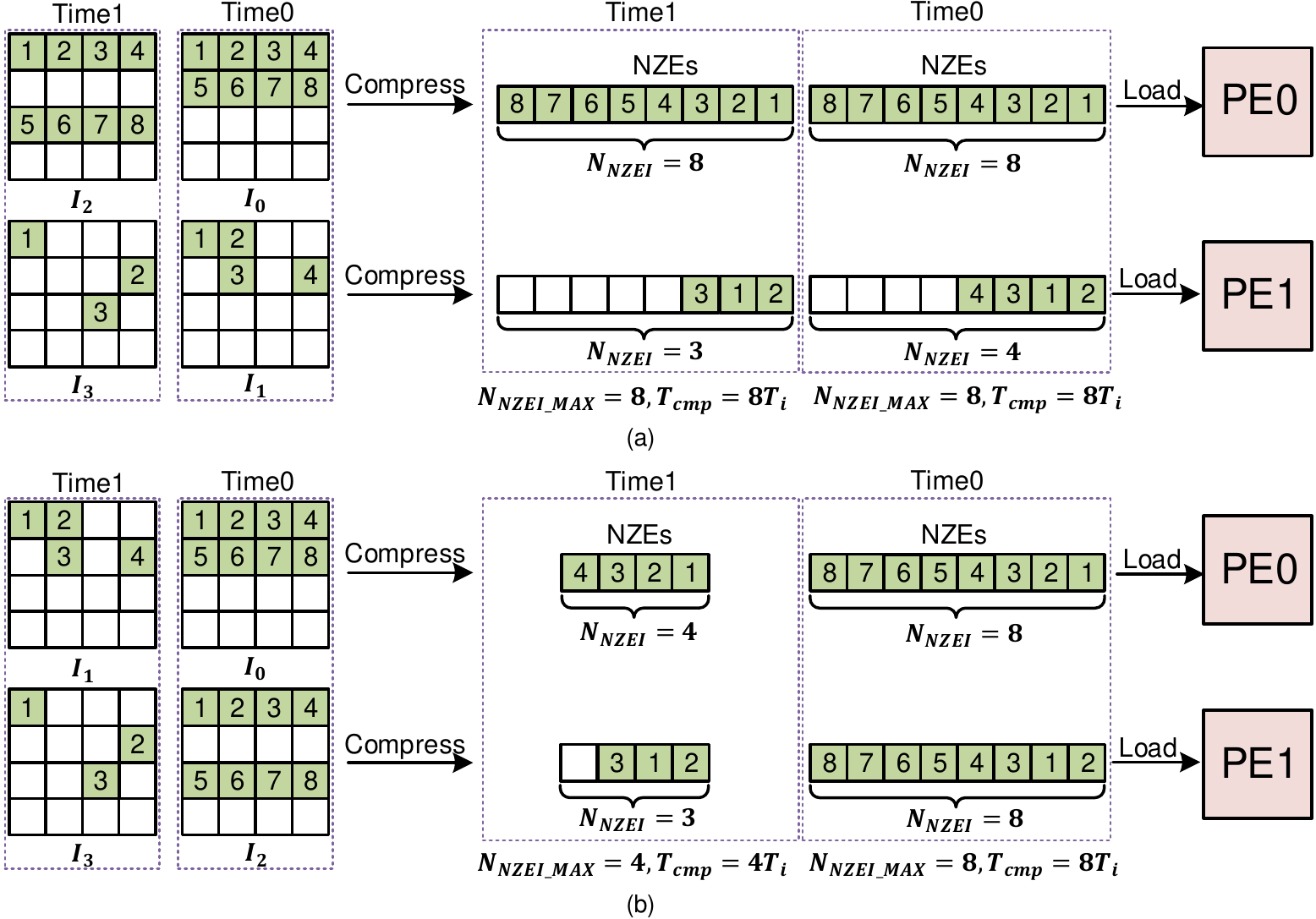}
	\caption{Imbalanced and balanced load of sparse IFMs in systolic array. (a)Imbalanced load.\iffalse The computing time of imbalanced workload is 16$T_i$ with 28\% PE idle rate.($T_i$: time for processing one IFM element).\fi(b)Balanced load.\iffalse The computing time of imbalanced workload is 12$T_i$, achieving $1.3\times$ speedup.\fi}
	\label{fig4}
\end{figure}

For sparsity of IFMs, there also exists many unevenly distributed zero elements after activation functions like ReLU, which still induce imbalanced workloads across PE array. Moreover, unlike weights, they are dynamically produced during execution and cannot be trained offline. Therefore, the sparsity rate of IFMs in different channels is unpredictable. As shown in Fig.\ref{fig4}(a),  IFMs of 4 channels are loaded into a $1\times2$ PE array in two time steps. Assuming the computation time of a single IFM element is $T_i$ and the numbers of NZEs in each IFM, $N_{NZEI}$, are 8, 4, 8, 3 respectively. If we load the compressed IFMs in order of $I_0$ and $I_1$, $I_2$ and $I_3$, the total computing time in $Time0$ and $Time1$ are both blocked to $8T_i$ for the largest numbers of NZEs, $N_{NZEW\_MAX}$, is 8, and $PE_1$ will be idle for $9T_i$ in total due to the imbalanced loads.\\ 
\indent\setlength{\parindent}{1em}Though there are some previous works of sparse CNN acceleration in systolic array, such as Swallow\cite{9026967}, FESA\cite{9218630} and SPOTS\cite{10.1145/3532863}, they suffer from inadequate use of sparsity of IFMs and weights due to imbalanced workloads. Thus, in this work, we aim to balance load of sparse weights and IFMs in systolic array architecture.

\section{Methodology}

To deal with imbalanced workload in systolic array, we applied load-balancing weight pruning for sparse weights and channel clustering for sparse IFMs, achieving great improvement of PE utilization and performance. For load-balancing weight pruning, we rank the elements of each kernel by importance, and then prune the less important elements to a certain sparsity ratio. For channel clustering of IFMs, we record the numbers of NZEs in each IFM, and accordingly rank the channel indexes. After that, IFMs can be accessed in the sequence of sorted channel indexes and gathered with approximate sparsity ratios. \\ 
\indent\setlength{\parindent}{1em}To fit the strategies of balancing workload in CONV layers, we directly compress IFMs and kernels for efficient sparsity utilization, and apply weight-oriented\cite{9130762} computing flow to coordinate with the compression format. Since FC layers, whose computation pattern is general matrix-vector multiplication(GEMV), only takes up a small proportion in the whole network, we map its computing flow as CONV pattern with outer product\cite{8327050} to reduce the control overhead. Additionally, for FC layers, we applied randomly pruning method\cite{7551397} to maximize sparsity of weights and channel clustering to balance workload of each column in weight matrices.

\subsection{Load-balancing Weight Pruning}

\begin{figure}[H]
	\centering
	\includegraphics[width=0.8\linewidth]{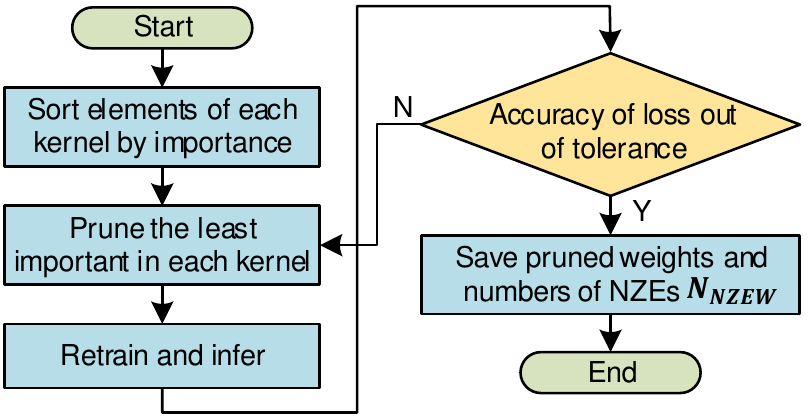}
	\caption{Load-balancing weight pruning flow.}
	\label{fig5}
\end{figure}

\begin{figure}[H]
	\centering
	\includegraphics[width=0.93\linewidth]{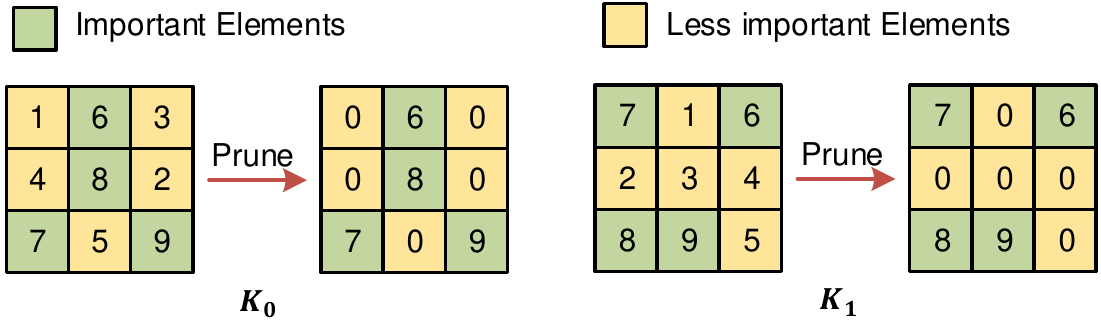}
	\caption{Example of load-balancing weight pruning. \iffalse(a)Pruning process of $3\times3$ kernel, achieving $2.25\times$ speedup. (b) Pruning process of $5\times5$ kernel with stride = 2 and 4 sub-tiles, achieving $2.08\times$ speedup.\fi}
	\label{fig6}
\end{figure}

CNNs contain many redundant weights, and with smaller absolute value of weight comes less importance. Thus, to balance computing loads across systolic array, we pruned each kernel by importance to set a number of less important elements zero and keep the sparsity ratios of all kernels at a certain value. An example of computing flows towards load-balancing weight pruning is shown in Fig.\ref{fig3}(b). Compared with load-imbalanced weight pruning in Fig.\ref{fig3}(a), the numbers of NZEs in $K_0$ and $K_1$ are both 4, eliminating idle PEs. Therefore we obtain the total computing time of $4T_w$ with $1.5\times$ speedup and greatly improve PE utilization rate.\\
\indent\setlength{\parindent}{1em}The specific flow of load-balancing weight pruning is shown in Fig.\ref{fig5}. Firstly, we sort elements of each kernel by importance and then set the least important element zero. After that, the previously pruned weights will be retrained and testified if the accuracy drops out of boundary. If not, we continue to prune the remaining elements and train, otherwise, we save the final pruned weights and record $N_{NZEW}$. Fig.\ref{fig6} shows a pruning example of two $3\times3$ kernels, $K_0$ and $K_1$. We set the 5 least important elements zero in each kernel. Though the distribution of zero elements is irregular, the workload of each kernel are balanced, therefore achieving $9/4 = 2.25\times$ speedup. 


\subsection{Channel Clustering}

\begin{figure*}
	\centering
	\includegraphics[width=0.75\linewidth]{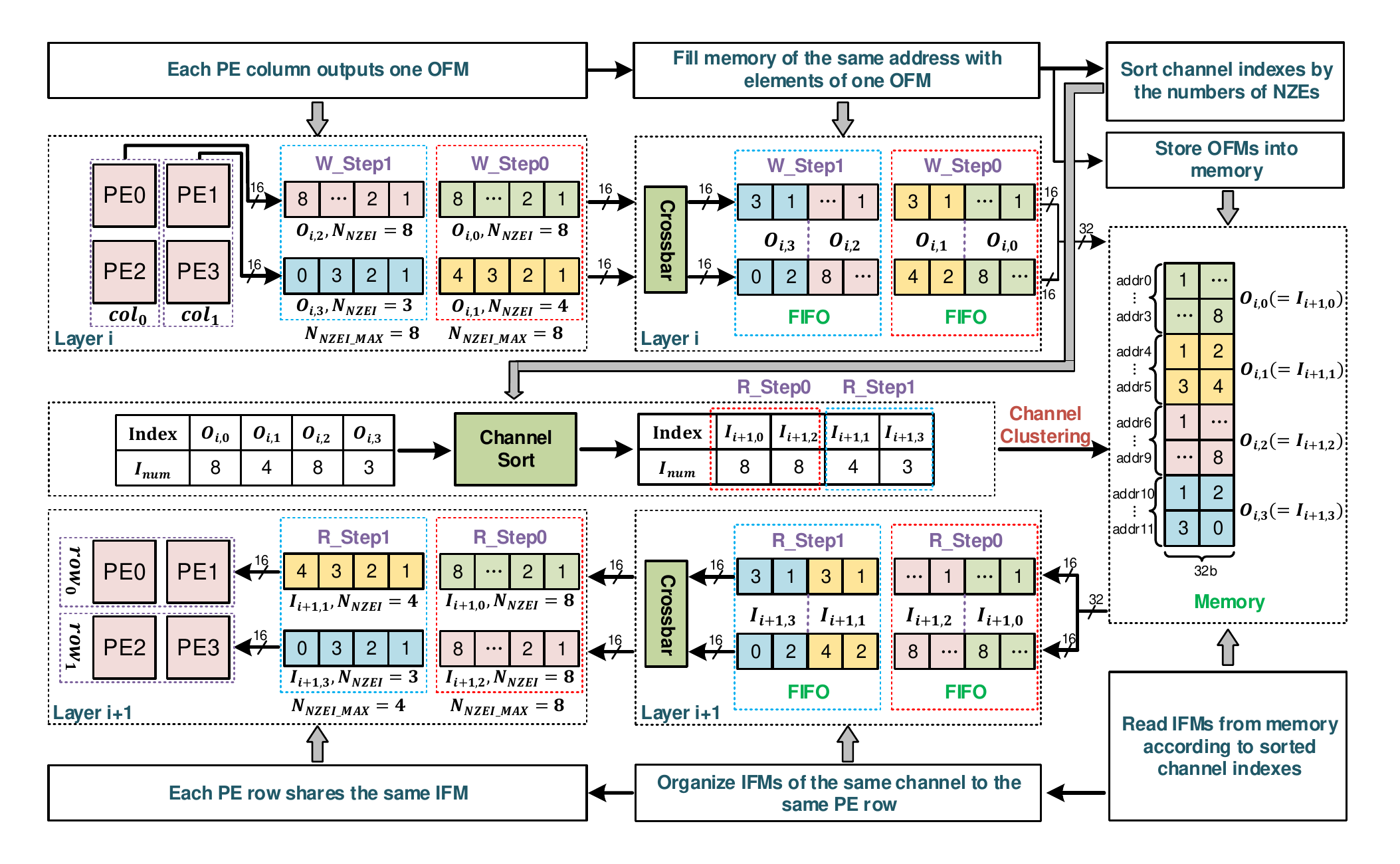}
	\caption{Process of channel clustering. $I_{i+1,0}$, $I_{i+1,2}$ and $I_{i+1,1}$, $I_{i+1,3}$ are gathered to compute respectively\iffalse, achieving $1.33\times$ PE utilization rate\fi.}
	\label{fig8}
\end{figure*}

Since IFMs are generated dynamically, they cannot be trained offline to balance IFM workload. Thus, we rank ICs by the numbers of NZEs in IFMs on hardware, so that channels with approximate sparsity ratios will be clustered to compute in PE array. An example of channel clustering is shown in Fig.\ref{fig4}(b). Compared with the flow without channel clustering in Fig.\ref{fig4}(a), we cluster $I_0$ and $I_2$, $I_1$ and $I_3$ respectively, and the total computing time in $Time0$ and $Time1$ reduces to $8T_i$ and $4T_i$ respectively, achieving $1.33\times$ speedup and greatly eliminating idle PEs. \\ 
\indent\setlength{\parindent}{1em}The specific flow of channel clustering is shown in Fig.\ref{fig8}. And from the perspective of dataflow, each PE row shares the same IFMs and each column outputs one OFM to save bandwidth and improve parallelism, which will be further discussed in Section V. $col_0$ and $col_1$ concurrently produce two channels of OFMs in each step during layer $i$, $O_{i,0}$ and $O_{i,1}$ of W\_step0, $O_{i,2}$ and $O_{i,3}$ of W\_step1. Meanwhile, we sort $O_{i,0}$\textasciitilde$O_{i,3}$ by the numbers of NZEs to cluster OFMs with approximate sparsity. Assuming memory of each address can accommodate outputs of 2 PE columns, if we directly write in the OFMs from W\_step stage, each input data of memory contains elements from two different channels, making it harder to read out a complete channel. However, due to the irregular access order of OFMs channel clustering and variation of the numbers of NZEs, $N_{NZEI}$, in each channel, it clearly consumes more time and energy to read out all required data based on this storage layout of OFMs. Thus, we fill memory of the same address with OFM elements of the same channel through crossbar and FIFO. In this way, channels can be efficiently accessed according to the sorting order of channel clustering,  $I_{i+1,0}$(OFM $O_{i,0}$ in previous layer $i$) and $I_{i+1,2}$ for R\_step0, $I_{i+1,1}$ and $I_{i+1,3}$ for R\_step1. But before sending IFMs to PE array, we need to organize IFMs of the same channel to the same PE row to fit the computing flow. Eventually, $row_0$ and $row_1$ receive IFMs with approximate sparsity ratios, which balances computing loads across PE array and reduces idle duration of each PE, achieving $1.33\times$ PE utilization rate with little hardware cost.  

\subsection{Sparsity Processing of CONV Layers}

\begin{figure}[H]
	\centering
	\includegraphics[width=0.7\linewidth]{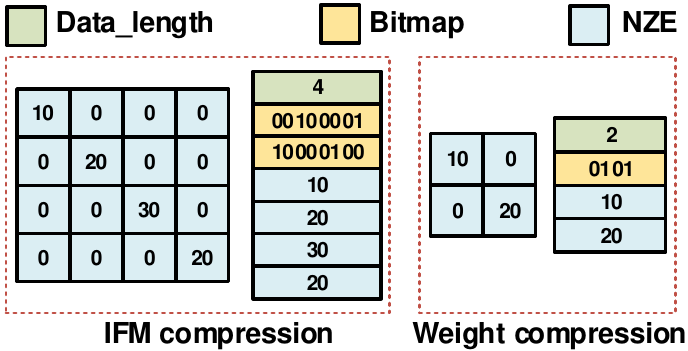}
	\caption{Compression patterns \iffalse of IFM and weight\fi for CONV layer.}
	\label{fig10}
\end{figure}

\begin{figure}[H]
	\centering
	\includegraphics[width=0.7\linewidth]{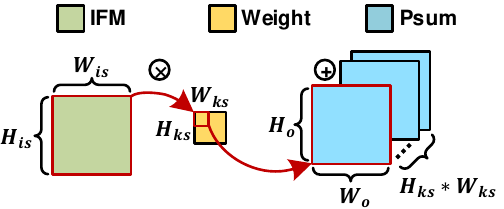}
	\caption{Weight-oriented dataflow.}
	\label{fig9}
\end{figure}

For CONV layer, we compress the whole kernel and IFM block to make full use of sparsity of IFMs and weights with bitmap compression format\cite{10.1145/3352460.3358291} as shown in Fig.\ref{fig10}. Each compressed data block contains the data\_length, bitmap and NZEs. The data\_length, $N_{NZEI}$ and $N_{NZEW}$, records the number of NZEs, bitmap records data zero flags(0 for zero, 1 for nonzero), and NZEs are represented as $I_{NZ}$ and $W_{NZ}$. To fit with the compression format, we applied weight-oriented\cite{9130762} computing flow, shown in Fig.\ref{fig9}. It takes one weight each time to compute with the whole IFM, repeatedly, whose results will be accumulated until the final results are obtained. 
To obtain OFMs in sparse processing, we first traverse IFMs then weights and decompress corresponding bitmap into location info, $(I_{row},I_{col})$ and $(W_{row},W_{col})$. Based on that, output location $(Psum_{row}(=I_{row}-W_{row}),Psum_{col}(=I_{col}-W_{col}))$ is calculated and set as invalid when out of boundary. Then according to the output location and edge size of OFM, $W_o$, we obtain buffer address, $Psum_{addr}(=Psum_{row}*W_o+Psum_{col})$, of accumulated partial sums(Psums) for the product of $I_{nz}$ and $W_{nz}$.

\begin{figure}[H]
	\centering
	\includegraphics[width=0.9\linewidth]{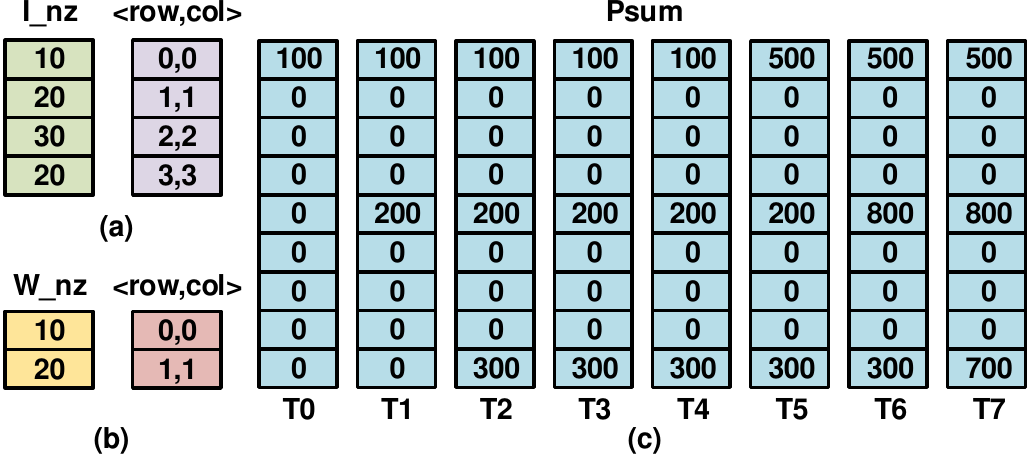}
	\caption{Sparse computing process of CONV layer.}
	\label{fig11}
\end{figure}
Fig.\ref{fig11} shows an example. The decompression of IFMs is shown in Fig.\ref{fig11}(a). Assuming  $W_{o}=3$, we have $I_{NZ}=\{10, 20,30,40\}$, whose location info are $(I_{row}$, $I_{col})=\{(0,0),(1,1),(2,2),(3,3)\}$. Fig.\ref{fig11}(b) shows the decompression of kernels. Assuming $W_{NZ}=\{10, 20\}$, we have location info, $(W_{row}$, $W_{col})=\{(0,0),(1,1)\}$. The computing flow is shown in Figure \ref{fig11}(c). Assuming the initial value of Psum is zero, at $T_0$, we fetch $I_{NZ} = 10$, $(I_{row}, I_{col})=(0,0)$, and $W_{NZ} = 10$, $(W_{row}, W_{col})=(0,0)$ and calculate out $Psum = 100$, and $Psum_{addr} = 0$, so value $100$ is accumulated in address $0$ of Psum buffer. Similarly at $T_1$, $T_2$, value $200$ and $300$ are accumulated in address $4$ and $9$ respectively. Since $Psum_{addr}$ is out of boundary at $T_3$ and $T_4$, operations are regarded as invalid. Finally at $T_5$, $T_6$ and $T_7$,  $400$, $600$, $400$ are accumulated with previous values in address $0$, $4$, $9$ respectively. The whole process takes $8$ cycles ,  $8\times$ faster than $64$ cycles of the original process.

\subsection{Sparse Processing of FC Layers}

\begin{figure}[H]
	\centering
	\includegraphics[width=0.52\linewidth]{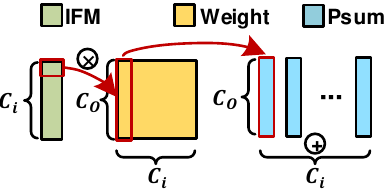}
	\caption{Process of outer product.}
	\label{fig12}
\end{figure}

\begin{figure}[H]
	\centering
	\includegraphics[width=0.7\linewidth]{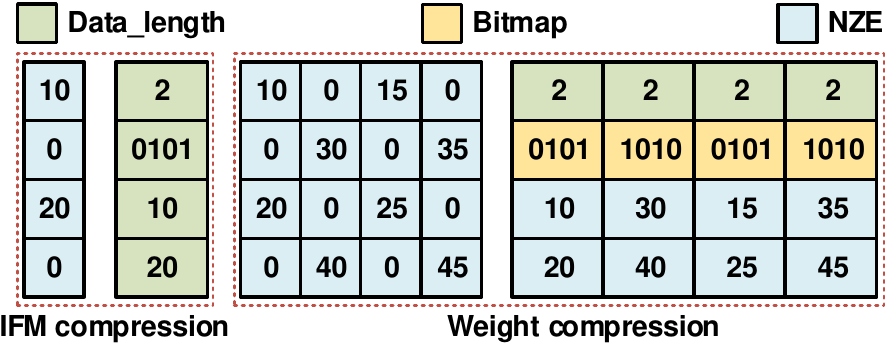}
	\caption{Compression patterns \iffalse of IFM and weight\fi for FC layer.}
	\label{fig13}
\end{figure}

Since FC layers mainly conduct GEMV and only take up a small proportion of CNN, we map the computing flow as CONV pattern to reduce control overhead. Therefore, we applied outer product\cite{8327050} to process FC layers, as shown in Fig.\ref{fig12}. It takes one element of IFM each time to compute with the corresponding column of weight matrices, repeatedly, whose results will be accumulated until all inputs are traversed. Correspondingly as in CONV layers, the temporary results of each input channel are accumulated to obtain a complete OFM. To efficiently exploit sparsity, IFMs and weights are all compressed by column as shown in Fig.\ref{fig13}. Since the computation is mainly GEMV, the total runtime is determined by the size of weight matrix. Thus, we perform random pruning\cite{7551397} to maximize weight sparsity. Moreover, for weight loads balancing of FC layers, channel clustering is applied like CONV layers by sorting the numbers of NZEs in each weight column, further improving performance.

The sparse process of FC layers is similar to CONV layers. First, decompress the location info of IFMs and weights. Then fetch nonzero IFM element, $I_{NZ}$ with location info of $I_{col}$, and nonzero weights of the $I_{col}$ column of weight matrix with location info of ($W_{row}$,$W_{col}$). Eventually, we further calculate the $Psum_{addr}$($=W_{row}$) of Psum Buffer for Psum accumulation. 

\section{Architecture }

To achieve high speedup and energy efficiency at low hardware cost and power consumption, we chose systolic array as mainstay. Then, how to efficiently fit in sparsity processing becomes the major problem. We equipped compressing module to store data in compressed format, and added channel clustering module to gather IFM computations with approximate sparsity ratio, balancing computing workload of IFMs. Accordingly, decompressing module is designed for IFM and weight inputs to decompress location info. Moreover, based on our dataflow, we surrounded the array with input, weight and output buffers to maximize data reuse and minimize DRAM access. The design of overall architecture aims to balance workload of sparse IFMs and weights in systolic array, collaborating with load-balancing weight pruning and channel clustering.

\subsection{Overview}

\begin{figure}[H]
	\centering
	\includegraphics[width=0.85\linewidth]{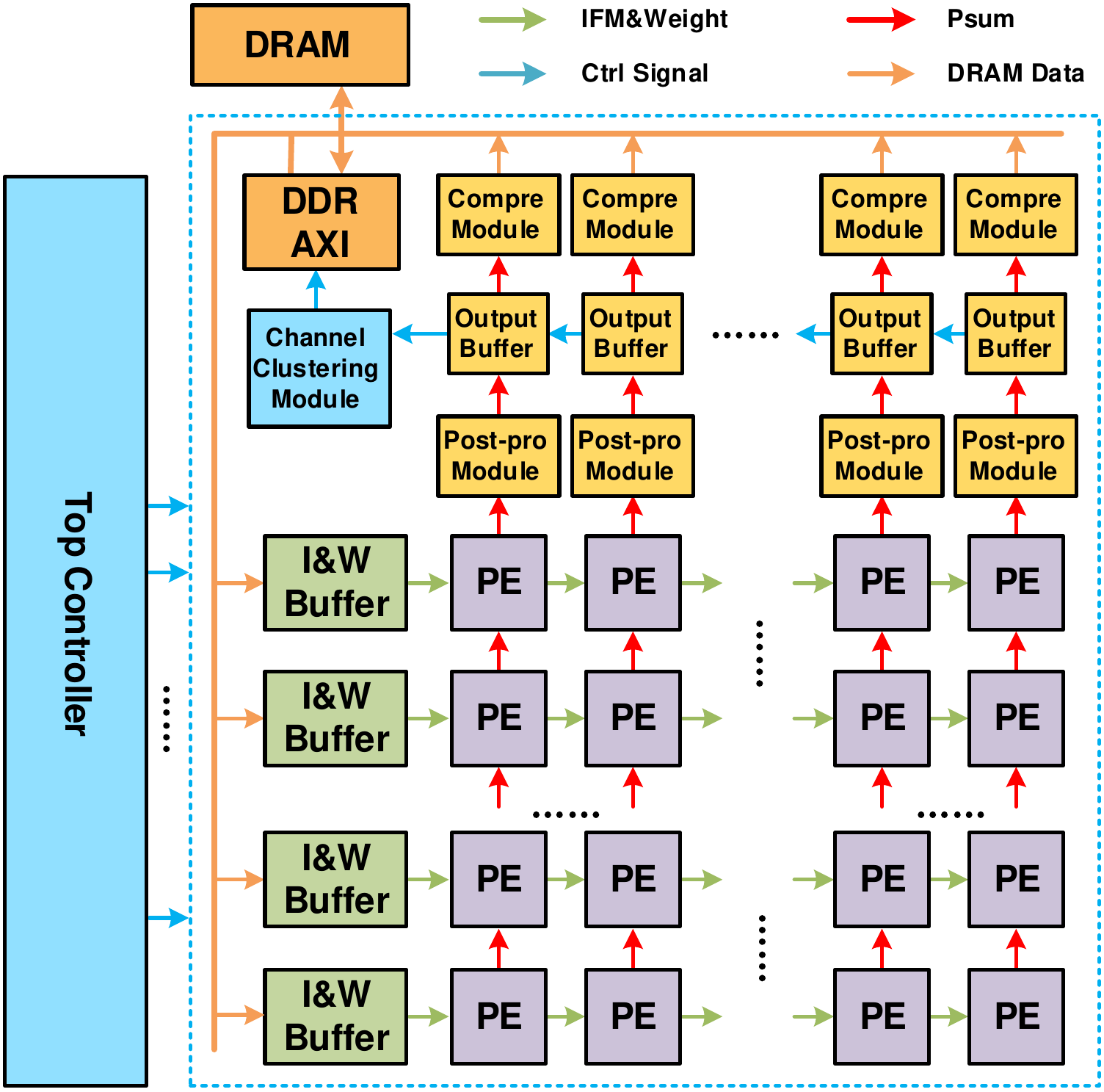}
	\caption{Top-level architecture.}
	\label{fig15}
\end{figure}

As shown in Fig.\ref{fig15}, the top level architecture mainly consists of input-weight buffers(I\&W buffer), a PE array, post-processing modules(Post-pro module), output buffers, compressing modules(Compre module), a top controller, a channel clustering module and DDR-AXI interface. Initially, we load pruned weights, input images and parameters directly into DRAMs, but in the following computing duration, the generated IFMs of each layers are stored in compressing format to reduce DRAM access. Then the compressed data are fetched and decompressed by I\&W buffers to obtain NZEs and location info for PE array. After parallel MAC operations by PEs, post-pro modules take in the outputs to perform ReLU, pooling, etc, whose results are buffered in the output buffers. Before storing back to DRAM, the OFMs are compressed by the compre modules. Meanwhile, channel clustering module sorts output channel indexes according to the number of NZEs in OFMs to gather them with approximate sparsity. We execute layer-wise, and OFMs of the last layer are classified to output the final results. Top controller masters the whole computing flow of network and DDR-AXI interface transports data on/off-chip. 
\subsection{On-chip Buffer}

To reduce energy consumption of data movement on/off chip, we need to buffer IFMs, weights and OFMs for data reusing. Thus, on-chip buffers mainly consist of I\&W buffers and output buffers; I\&W buffer consists of two BRAMs for IFMs and weights, two BRAMs for location info and a decompressing unit for bitmaps, and output buffer consists of two BRAMs for OFMs. To avoid performance from data loading, we utilize each group of two BRAMs as Ping-Pong buffers, one for DRAM access and the other for PE transporting. Inputs of each row of PE array are supplied by one I\&W buffer; outputs of each column are collected by one output buffer. Additionally, to support different reuse strategies, storage can be re-allocated for IFMs and weights.


\subsection{PE Array}
\begin{figure}[H]
	\centering
	\includegraphics[width=0.7\linewidth]{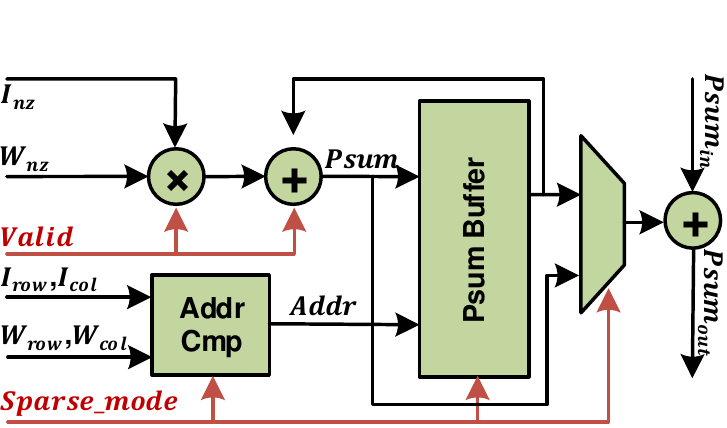}
	\caption{PE unit.}
	\label{fig16}
\end{figure}

The PE array, in charge of MACs, is composed of $N_{PE} \times N_{PE}$ PEs, among which PEs of the same row share IFMs of one IC and PEs of the same column process OFMs of one OC. Each PE contains a MAC unit, an address computing unit and a Psum buffer, as shown in Fig.\ref{fig16}. During execution, we multiply IFMs with weights and calculate location info. Results are accumulated with temporarily stored Psums of corresponding address in Psum Buffer. When all ICs of this output block are finished, we pause the computation, accumulate across PEs, and finally send the results to post-pro module. To reduce power consumption, the MAC unit will be gated when the IFM/weight is zero or the location info is invalid. Moreover, in the case of low sparsity, we switch the PE array to dense mode and accumulate the products of IFM and weight with adjacent PE, which can shield the address calculation unit and Psum buffer, thereby reducing the power consumption of PEs.

\subsection{Channel Clustering Module}
To balance load of sparse IFMs in PE array, we rank ICs by the numbers of NZEs in IFMs, so that channels with approximate sparsity ratios will be gathered in channel clustering module, which consists of a channel index buffer, an NZEs number buffer, a ranking unit, a crossbar and a group of FIFOs. Since only $N_{PE}$ OFMs are produced each time, we divide and rule, ranking each group of OFMs separately. According to the numbers of NZEs stored in NZEs number buffer, we sort the channel indexes with Merge Sort algorithm in ranking unit and the results are stored in channel index buffer. Since the IFM access of DRAM becomes irregular after channel clustering, we utilized a crossbar and a group of FIFOs to rearrange the OFMs layout, allowing efficient random access of OFMs. When processing the next layer, IFMs(OFMs in previous layer) are read in based on ranking indexes and gathered with approximate sparsity ratios to further balance IFM computing load in PE array.

\section{Dataflow}

Since IFMs and weights cannot be all stored on-chip, we partitioned IFMs into independent sub-tiles to decrease on-chip storage requirement. Additionally, for bandwidth saving and DRAM access reducing, we introduced several data reuse strategies and designed Adaptive Dataflow Configuration to decide a more suitable dataflow of each layer based on the storage ratios of weights and IFMs. Finally, a mapping algorithm is designed to map various networks on our architecture, based on provided network parameters.

\subsection{IFM \& Weight Partition}

For limited on-chip buffer and PE array size, we need to partition IFMs	into independent sub-tiles in terms of feature map edge size and input/output channel. Based on CONV, there’s overlap between sub-tiles; to reduce overlapping, we prefer square shape partition. Considering the resources of Psum storage, the size of feature map sub-tiles is set no larger than $N_{is} \times N_{is}$, and the numbers of IFM and OFM tiles on row/column dimension are $T_{ifm\_row}$, $T_{ifm\_col}$ and $T_{ofm\_row}$, $T_{ofm\_col}$ respectively. For limited PE array size, we only process $N_{PE}$ input and output channels concurrently, with tiling numbers of $T_{ic}$ and $T_{oc}$. To obtain complete OFMs, all the input sub-tiles will convolve with kernels in turn. 

\subsection{Reuse Strategy}

To avoid same IFMs and weights being read repeatedly from DRAM and reduce Psum storage, we attempt to consume data as soon as possible and write back only the final results to DRAM. Therefore, we apply output stationary strategy\cite{7551407}, taking IFMs and weights of different input channels on-chip to compute and accumulate in the same column to obtain the OFM of this output channel. Additionally, based on the intrinsic property of systolic architecture, the same PE row shares one IFM, which actualizes input stationary strategy\cite{7551407} and improves bandwidth utilization rate. As for a single PE, we introduce weight stationary strategy\cite{7551407}, temporally reusing weights when the whole IFM streams through. In this way, we maximize data reuse by fully applying input, weight and output stationary strategies, minimizing data movement consumption. For FC layers, since there exists no weight reusing in GEMV operation, the computing time is mainly determined by weight loading. Limited by the bandwidth, we only applied one PE column to process FC layers for energy reduction of the whole system.

\subsection{Adaptive Dataflow Configuration}
\begin{figure}[H]
	\centering
	\includegraphics[width=0.97\linewidth]{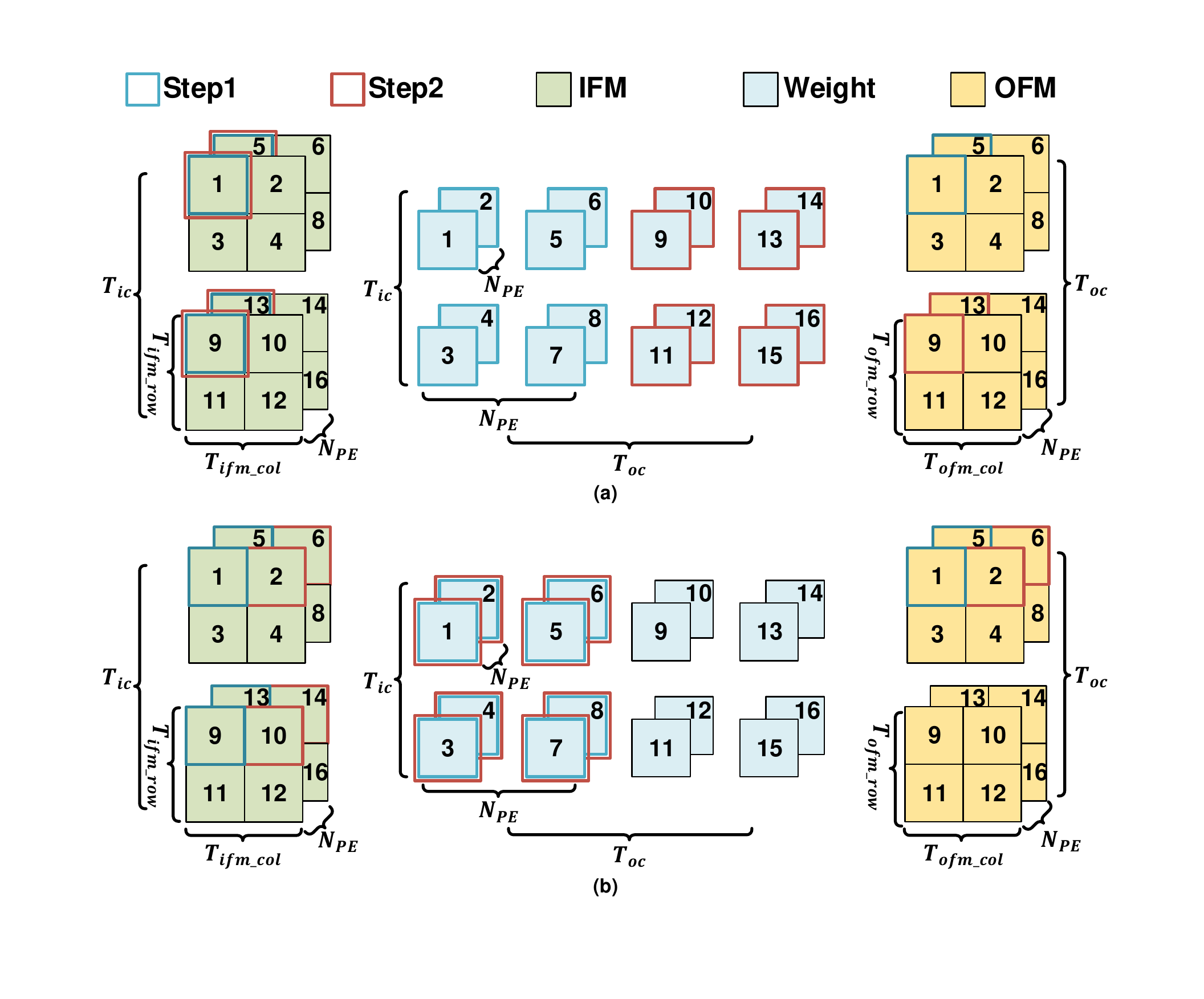}
	\caption{Adaptive dataflow configuration. (a)Reuse-IFM-First(RIF). (b)Reuse-Weight-First(RWF). }
	\label{fig17}
\end{figure}

After data partition, OFMs are tiled in edge size and channel dimension. Thus, the dataflow of OFMs faces two choices: channel first or edge first. To further reduce DRAM access, we discussed the impact of two dataflows on DRAM access as shown in Fig.\ref{fig17}. For channel first, we calculate $I_1$, $I_5$, $I_9$ and $I_{13}$ with $K_0$\textasciitilde$K_7$ to obtain $O_1$ and $O_5$ in Step0, and then we station IFMs and traverse weights $K_8$\textasciitilde$K_{15}$ to obtain $O_9$ and $O_{13}$, as shown in Fig.\ref{fig17}(a); this dataflow reuses IFMs and repeatedly access weights $T_{ifm\_col} \times T_{ifm\_row}$ times, thus called “Reuse-IFM-First(RIF)”. Assuming the storages of IFMs and weights are $I_{mem}$ and $W_{mem}$ respectively, the total DRAM access, $D_{mem}$, is $W_{mem} \times T_{ifm\_row}\times T_{ifm\_row}+I_{mem}$. \\
On the contrary, if we calculate out the sub-tiles $O_1$, $O_5$ first and then $O_2$, $O_6$ as shown in Fig.\ref{fig17}(b), we will station weights and repeatedly access IFMs $T_{oc}$ times, which is called “Reuse-Weight-First(RWF)”. Thus, $D_{mem}$ is $I_{mem} \times T_{oc}+W_{mem}$. Additionally, when all weights can be stored on-chip, we can station IFMs without loading weight repeatedly. Therefore, $D_{mem}$ is $I_{mem} + W_{mem} $. As DRAM access varies based on different dataflows, to further reduce DRAM access, we applied Adaptive Dataflow Configuration to decide a more suitable reuse strategy, “RIF” or “RWF”, of each layer based on the storage ratios of weights and IFMs. For FC layers, since IFMs can be totally stored on-chip and weights require no reusing, $D_{mem}$ is $I_{mem} + W_{mem} $.

Tab.\ref{tab:tab2} shows an example of Adaptive Dataflow Configuration on different layers in ResNet-50. Assuming $N_{is}$ is $7\times7$ and $N_{PE}$ is 32, we will apply different reuse strategies on layer-3, layer-15 and layer-48 respectively. In layer-3, $T_{ifm\_row}$ and $T_{ifm\_col}$ are both 8; $T_{ic}$ and $T_{oc}$ are both 2. Since weights can entirely be loaded on-chip for its small quantity, we apply RIF strategy to avoid repeated access of weights. In layer-15, $T_{ifm\_row}$, $T_{ifm\_col}$, $T_{ic}$ and $T_{oc}$ are all 4. RWF strategy reduces $4.49\times$ DRAM access than RIP strategy. In layer-48, when $T_{ifm\_row}$ and $T_{ifm\_col}$ are both 1, $T_{ic}$ and $T_{oc}$ are both 16, IFMs are reused prior, achieving $1.16\times$ reduction of DRAM access. 

\begin{table}[H]\centering
	\caption{Reuse Strategies in Different Cases}
	\label{tab:tab2}
	\begin{tabular}{|c|c|c|c|c|c|}
		\hline
		Layer    & \begin{tabular}[c]{@{}c@{}}$T_{ic}$, $T_{oc}$,  \\ $T_{ifm\_row}$,\\  $T_{ifm\_col}$\end{tabular} & $I_{mem}$ & $W_{mem}$ & \begin{tabular}[c]{@{}c@{}}RWF\\  $D_{mem}$  \\ \end{tabular} & \begin{tabular}[c]{@{}c@{}}RIF\\ $D_{mem}$   \\ \end{tabular} \\ \hline
		Layer-3  & 2,2,8,8                                                                                           & 196K      & 36K       & /                                                             & 232K                                                             \\ \hline
		Layer-15 & 4,4,4,4                                                                                           & 98K       & 144K      & 536K                                                        & 2.35M                                                         \\ \hline
		Layer-48 & 16,16,1,1                                                                                         & 24.5K     & 1.99M     & 2.63M                                                         & 2.27M                                                         \\ \hline
	\end{tabular}
\end{table}

\subsection{Mapping Algorithm }

Considering dataflow and partition of IFM and weight, to map various networks on Sense, we designed a network mapping algorithm, obtaining architecture configuration parameters through feeding network structure parameters. With configuration parameters loaded, the controller can master the whole network computing flow. \\ 
\indent\setlength{\parindent}{1em}Computing flow description is shown in Tab.\ref{tab:tab4}. The 1st and 4th row decides the dataflow according to reuse strategy; for "RIF", we finish the OFM computing of all OCs for one output tile before switching to the next output tile; otherwise, we finish computing of all output tiles for one OC before switching to the next OC. The 2nd and 3rd row describe the partition of OFM, with row number of $T_{ofm\_row}$ and column number of $T_{ofm\_col}$. The 5th row drives the accumulation of input channels to get the final Psums. The 6th row and the 7th represent the numbers of NZEs in one tile of IFM and one kernel; $N_{NZEW\_MAX}$ is loaded as a parameter because the weights from training are fixed and $N_{NZEI\_MAX}$ is gained during data loading. The 8th row means the PE is performing MAC operation with NZEs of IFM and weight.

	\begin{table}[H]\centering
	\caption{Computing Flow}
	\label{tab:tab4}
	\begin{tabular}{lll}
		\toprule
		\textbf{Input}: parameters of architecture configuration; IFMs; weights;\\
		\textbf{Output}: OFMs \\
		\midrule
		When RIF, $T_{oc\_outter} = 1, T_{oc\_inner} = Toc; otherwise,$\\
		 $T_{oc\_outter} = Toc, T_{oc\_inner} = 1$\\
		\textbf{1}\hspace{0.2cm}$for(a=0;a<T_{oc\_outter};a=a+1)$	\\
		\textbf{2}\hspace{0.4cm}$ for(b=0;b<T_{ifm\_row};b=b+1)$	\\
		\textbf{3}\hspace{0.6cm}$  for(c=0;c<T_{ifm\_col};c=c+1)$	\\
		\textbf{4}\hspace{0.8cm}$   for(d=0;f<T_{oc\_inner};d=d+1)	$\\
		\textbf{5}\hspace{1cm}$      for(e=0;e<T_{ic};e=e+1)$ \\
		\textbf{6}\hspace{1.2cm}$     for(f=0;f<N_{NZEW\_MAX};f=f+1)$	\\
		\textbf{7}\hspace{1.4cm}$ 	    for(g=0;g<N_{NZEI\_MAX};g=g+1)$	\\
		\textbf{8}\hspace{1.6cm}$	     Psum+=W_{NZ}\times I_{NZ}$\\
		
		\bottomrule
	\end{tabular}
\end{table}

\section{Experiments}

\subsection{Implementation}

We implement Sense in Verilog, and conduct functional simulation on Vivado 2021.1. Implementation is run on ZynqZCU102 at 200MHz and we obtain the reports of resource and power from Vivado, as shown in Tab.\ref{tab:tab5}. For DRAM power evaluation, we simulate on CACTI\cite{10.1145/3085572} platform according to the number of DRAM accesses. \\
\indent\setlength{\parindent}{1em}The size of PE array is $32\times32$, achieving a peak throughput of $1024\times0.2GHz = 204.8GMAC/s$. IFMs and weights are encoded in 16bit and the Psum is truncated to 16bit for storage reduction with little accuracy loss through training. To balance data reusing efficiency and resource overhead, we set the IFM tiling unit as $7\times7$. Thus, the buffers in PE array and compress modules are implemented with $64\times16b$ LUT RAM to minimize on-chip storage. Seen from the resource utilization breakdown shown in Tab.\ref{tab:tab5}, LUTs and DSPs are mainly used to construct the PE array for MAC operation and the majority of BRAMs are consumed in I\&W buffers for data reusing. Besides, the power and resource of channel clustering module only takes up a small proportion, indicating low cost for sparse processing in our architecture. \\

\begin{table}[H]\centering
	\caption{Resource and Power Proportion\iffalse of Each Module\fi}
	\label{tab:tab5}
	\begin{tabular}{|c|c|c|c|c|}
		\hline
		Module                                                           & DSP                                                        & LUT                                                        & BRAM                                                      & POWER                                                     \\ \hline
		\begin{tabular}[c]{@{}c@{}}I\&W    \\ Buffer\end{tabular}        & 0                                                          & \begin{tabular}[c]{@{}c@{}}8K   \\ (1.4\%)\end{tabular}    & \begin{tabular}[c]{@{}c@{}}320\\ (17.5\%)\end{tabular}    & \begin{tabular}[c]{@{}c@{}}1.08W   \\ (10\%)\end{tabular} \\ \hline
		\begin{tabular}[c]{@{}c@{}}PE\\ Array\end{tabular}               & \begin{tabular}[c]{@{}c@{}}1024   \\ (40.6\%)\end{tabular} & \begin{tabular}[c]{@{}c@{}}241K  \\ (40\%)\end{tabular}    & 0                                                         & \begin{tabular}[c]{@{}c@{}}5.62W  \\ (52\%)\end{tabular}  \\ \hline
		\begin{tabular}[c]{@{}c@{}}Post\_pro \\ Module\end{tabular}      & 0                                                          & \begin{tabular}[c]{@{}c@{}}11K  \\ (1.8\%)\end{tabular}    & \begin{tabular}[c]{@{}c@{}}32  \\ (1.8\%)\end{tabular}    & \begin{tabular}[c]{@{}c@{}}0.1W \\ (1\%)\end{tabular}     \\ \hline
		\begin{tabular}[c]{@{}c@{}}Output   \\ Buffer\end{tabular}       & 0                                                          & \begin{tabular}[c]{@{}c@{}}4K   \\ (0.7\%)\end{tabular}    & \begin{tabular}[c]{@{}c@{}}64  \\ (3.5\%)\end{tabular}    & \begin{tabular}[c]{@{}c@{}}0.1W \\ (1\%)\end{tabular}     \\ \hline
		\begin{tabular}[c]{@{}c@{}}Compre    \\ Module\end{tabular}      & 0                                                          & \begin{tabular}[c]{@{}c@{}}13K  \\ (2.2\%)\end{tabular}    & 0                                                         & \begin{tabular}[c]{@{}c@{}}0.1W  \\ (1\%)\end{tabular}    \\ \hline
		\begin{tabular}[c]{@{}c@{}}Channel    \\ Clustering\end{tabular} & 0                                                          & \begin{tabular}[c]{@{}c@{}}6K   \\ (1.1\%)\end{tabular}    & \begin{tabular}[c]{@{}c@{}}35   \\ (1.9\%)\end{tabular}   & \begin{tabular}[c]{@{}c@{}}0.3W  \\ (3\%)\end{tabular}    \\ \hline
		DDR-AXI                                                          & \begin{tabular}[c]{@{}c@{}}10   \\ (0.3\%)\end{tabular}    & \begin{tabular}[c]{@{}c@{}}37K   \\ (6.1\%)\end{tabular}   & \begin{tabular}[c]{@{}c@{}}51   \\ (2.8\%)\end{tabular}   & \begin{tabular}[c]{@{}c@{}}2.27W   \\ (21\%)\end{tabular} \\ \hline
		Controller                                                       & \begin{tabular}[c]{@{}c@{}}34   \\ (1.3\%)\end{tabular}    & \begin{tabular}[c]{@{}c@{}}28K   \\ (4.7\%)\end{tabular}   & 0                                                         & \begin{tabular}[c]{@{}c@{}}1.19W   \\ (11\%)\end{tabular} \\ \hline
		Total                                                            & \begin{tabular}[c]{@{}c@{}}1061   \\ (42.2\%)\end{tabular} & \begin{tabular}[c]{@{}c@{}}348K   \\ (58\%)\end{tabular} & \begin{tabular}[c]{@{}c@{}}502   \\ (27.5\%)\end{tabular} & \begin{tabular}[c]{@{}c@{}}10.8W   \\ (100\%)\end{tabular} \\ \hline
	\end{tabular}
\end{table}

To prove the superiority of Sense, we set some baseline accelerators as follows:\\		
\indent\setlength{\parindent}{1em}1) Swallow\cite{9026967} is a systolic architecture that fully utilizes sparsity of both weights and IFMs in CONV and FC layers, and applied a sparse-aware dataflow to optimize data reusing and lower DRAM access. By comparison, Sense achieves higher speedup through load-balancing weight pruning and channel clustering. Additionally, we further lower DRAM access by applying Adaptive Dataflow Configuration.\newline
\indent\setlength{\parindent}{1em}2) FESA\cite{9218630} proposes a software-hardware co-design to address the problem of no-load PEs for sparse networks and reduce the kernel pattern numbers of irregularly pruned networks, greatly improving PE utilization and performance. But it has not yet implemented on complex datasets like ImageNet, for its relatively strict constraints of pruning method. By comparison, we obtain similar performance improvement based on weight sparsity, which is also verified on ImageNet. Moreover, by exploiting sparsity of both IFMs and weights, we achieve $1.95\times$\textasciitilde$2.5\times$ performance improvement.\newline
\indent\setlength{\parindent}{1em}3) SPOTS\cite{10.1145/3532863} is a systolic-array-based accelerator for sparse CNNs by building a hardware unit to perform Im2Col transformation of IFMs coupled with GEMM unit. Moreover, a group-wise pruning method is designed to avoid the potential load imbalance caused by irregularity of sparse weights. By comparison, Sense can achieve higher sparsity of pruned weights with a more coarse-grained pruning method.\\
\indent\setlength{\parindent}{1em}4) Zhu $et$ $al$\cite{9130762} and Lu $et$ $al$\cite{8735526} are sparsity-aware CNN accelerators based on FPGA. Unlike systolic dataflow, they process sparse computing in independent PE groups with large logic and BRAM resource overhead. By comparison, we achieve approximate performance with less resource consumption.\newline
As for benchmarks, we conduct inference on AlexNet\cite{10.1145/3065386}, VGG-16\cite{simonyan2015deep}, ResNet\cite{7780459}, GoogleNet\cite{7298594} and compare our design with each baseline.

\subsection{Performance}

\begin{figure}[H]
	\centering
	\includegraphics[width=0.99\linewidth]{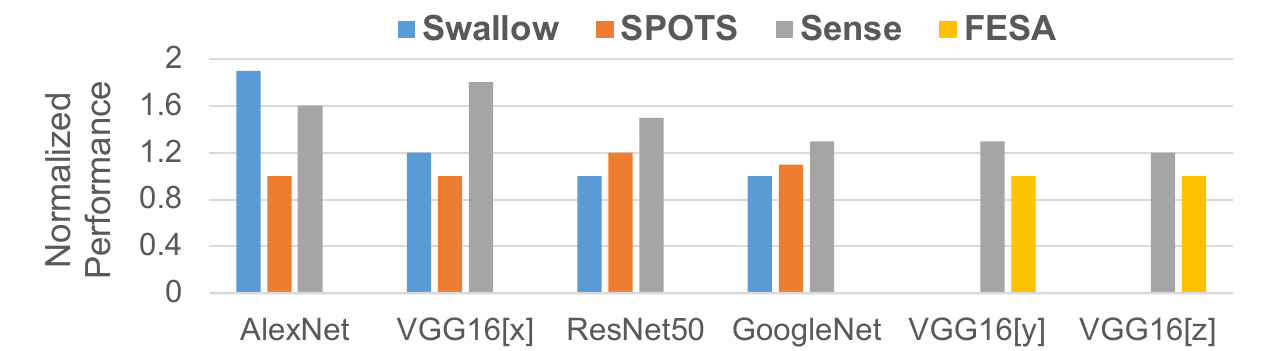}
	\caption{Performance comparison \iffalse of Sense\fi with Swallow, FESA and SPOTS in terms of sparse weights. \iffalse Sense achieves $0.84\times$\textasciitilde$1.53\times$, $1.2\times$\textasciitilde$1.3\times$ and $1.17\times$\textasciitilde$1.8\times$ speedup compared with Swallow, FESA and SPOTS respectively.\fi}
	\label{fig18}
\end{figure}

To show the advantages of sparse processing methods of Sense, we compared our design against Swallow, FESA and SPOTS on performance respectively, achieving $1\times$\textasciitilde$2.25\times$, $1.95\times$\textasciitilde$2.5\times$ and $1.17\times$\textasciitilde$2.37\times$ speedup as shown in Fig.\ref{fig20}. Tab.\ref{tab:tab6} displays the sparsity ratios of IFMs and weights in CONV and FC layers. In AlexNet, VGG-16[x](ImageNet dataset), ResNet and GoogleNet, we cut down the first 50\% small elements of each kernel in CONV layers with load-balancing weight pruning and set the first 80\% small elements of whole weight matrix in FC layers with randomly weight pruning\cite{7551397}. And In VGG-16[x](Cifar-10 dataset) and VGG-16[z](Cifar-100 dataset), we cut 78\% elements of each kernel in CONV layers and 80\% elements in FC layers. The columns of Top 1 and Top 5 accuracy shows the inference accuracy of each CNN and its corresponding accuracy loss in the brackets. The training structures refer to Pytorch\cite{Pytorch}. \\
\indent\setlength{\parindent}{1em}From the perspective of weight sparsity,  in Swallow, the distribution of zero elements in each kernel is irregular and the computing duration is determined by the least sparse kernel, causing unstable speedup. Although our pruning method may lead to relatively lower sparsity compared with Swallow due to the constraints, we managed to keep the sparsity ratio of each kernel at a certain value, which further balances PE loads across array and maintain a rather stable speedup, achieving higher performance. Thus, comparing with Swallow, we obtain $1.53\times$, $1.49\times$, $1.33\times$ speedup on VGG-16, ResNet-50 and GoogleNet respectively as shown in Fig.\ref{fig18}. However, since the weight sparsity ratio of AlexNet CONV layers in Swallow is 31.8\% higher than that of Sense, the speedup brought by CONV computing reduction is slightly higher than weight load balancing. Thus, Sense is $1.19\times$ slower than Swallow on AlexNet in total. \\
\indent\setlength{\parindent}{1em}As for the pruning method in FESA, it balances PE loads through reducing distribution patterns of zero elements in each kernel, but comes with strict constraints. Sense achieves $1.2\times$\textasciitilde$1.3\times$ performance improvement and loosens the constraints of FESA to implement more complex datasets like ImageNet. \\
\indent\setlength{\parindent}{1em}For group-wise pruning in SPOTS, it is more general for sparse CNNs, but sacrifices sparsity of weights due to the fine-grained pruning method. The sparsity of weights in Sense is $1\times$\textasciitilde$2.02\times$ higher than SPOTS and we achieve $1.17\times$\textasciitilde$1.8\times$ performance compared with SPOTS.


\begin{table*}\centering
	\caption{IFM and Weight Sparsity Ratios and Accuracy Loss of Each CNN}
	\label{tab:tab6}
	\begin{threeparttable} 
		\begin{tabular}{|c|c|c|c|c|c|c|c|}
			\hline
			\multirow{2}{*}{Accelerator}   & \multirow{2}{*}{CNN   type}   & \multirow{2}{*}{$W_{CONV}$}  & \multirow{2}{*}{$W_{FC}$}    & \multirow{2}{*}{$IFM_{CONV}$} & \multirow{2}{*}{$IFM_{FC}$}  & Top 1    & Top 5    \\ 
			& 						&				&							&				&			&	Accuracy	&	Accuracy \\ \hline
			\multirow{4}{*}{Swallow} & Alexnet      & 87.4\% & 81.1\% & 19.0\%    & 71.8\% & /                & 81.3\%(-0.3\%)   \\ \cline{2-8} 
			& VGG-16{[}x{]} & 62.8\% & 82.5\% & 39.5\%  & 33.4\% & /                & 88.2\%(-0.1\%)   \\ \cline{2-8} 
			& ResNet-50     & 46.9\% & 91.5\% & 46.2\%  & 22.0\% & /                & 90.6\%(-0.0\%)   \\ \cline{2-8} 
			& GoogleNet    & 58.1\% & 90.7\% & 44.0\%  & 22.9\% & /                & 91.0\%(-1.0\%)   \\ \hline
			\multirow{2}{*}{FESA}    & VGG-16{[}y{]} & 82.5\% & /      & /       & /      & 91.9\%(-0.9\%)   & /                \\ \cline{2-8} 
			& VGG-16{[}z{]} & 80.6\% & /      & /       & /      & 72.4\%(-0.4\%)   & 91.9\%(+0.5\%)                \\ \hline
			\multirow{4}{*}{SPOTS}   & Alexnet      & 56.8\% & /      & 34.2\%  & /      & 55.3\%(-1.5\%)   & 78.6\%(-1.3\%)   \\ \cline{2-8} 
			& VGG-16{[}x{]} & 27.5\% & /      & 49.7\%  & /      & 67.2\%(-1.1\%)   & 88.2\%(-0.2\%)   \\ \cline{2-8} 
			& ResNet-50     & 31.5\% & /      & 71.1\%  & /      & 69.7\%(-3.0\%)     & 89.3\%(-1.4\%)   \\ \cline{2-8} 
			& GoogleNet    & 25.1\% & /      & 41.2\%  & /      & 66.2\%(-2.7\%)   & 87.6\%(-1.5\%)   \\ \hline
			\multirow{6}{*}{Sense}   & Alexnet      & 55.6\% & 80.0\% & 35.8\%  & 76.3\% & 55.4\%(-0.8\%)   & 77.9\%(-0.4\%)   \\ \cline{2-8} 
			& VGG-16{[}x{]} & 55.6\% & 80.0\% & 49.2\%  & 83.2\% & 70.7\%(-0.9\%)     & 90.0\%(-0.4\%)   \\ \cline{2-8} 
			& ResNet-50     & 34.4\% & 80.0\% & 46.5\%  & 70.5\% & 75.4\%(-0.7\%)   & 92.6\%(-0.3\%)   \\ \cline{2-8} 
			& GoogleNet    & 33.5\% & 80.0\% & 34.7\%  & 60.2\% & 70.2\%(-0.8\%)     & 90.3\%(-2.2\%)   \\ \cline{2-8} 
			& VGG-16{[}y{]} & 77.8\% & 80.0\% & 43.6\%  & 47.1\% & 91.6\%(-0.5\%)     & /                \\ \cline{2-8} 
			& VGG-16{[}z{]} & 77.8\% & 80.0\% & 57.8\%  & 63.1\% & 91.8\%(-0.3\%)   & 99.8\%(-0.0\%)   \\ \hline
		\end{tabular}
		\begin{tablenotes}    
			\footnotesize               
			\item[1] VGG-16[x],VGG-16[y] and VGG-16[z] are verified on dataset of  ImageNet, Cifar-10 and Cifar-100 respectively. 
		\end{tablenotes}            
	\end{threeparttable}
	
\end{table*}

When it comes to IFMs, Swallow faces the problem of imbalanced load; FESA lacks of IFM sparsity handling; and SPOTS can only exploit the sparsity of IFMs in those rows with all zeros. To improve sparsity utilization of IFMs, we apply channel clustering and gather IFMs with approximate sparsity ratios. This treatment further balances PE loads and improve performance. Thus, we obtain $1.1\times$\textasciitilde$1.5\times$, $1.5\times$\textasciitilde$2.1\times$ and $1\times$\textasciitilde$1.35\times$ speedup compared with Swallow, FESA and SPOTS respectively. 

\begin{figure}[H]
	\centering
	\includegraphics[width=0.99\linewidth]{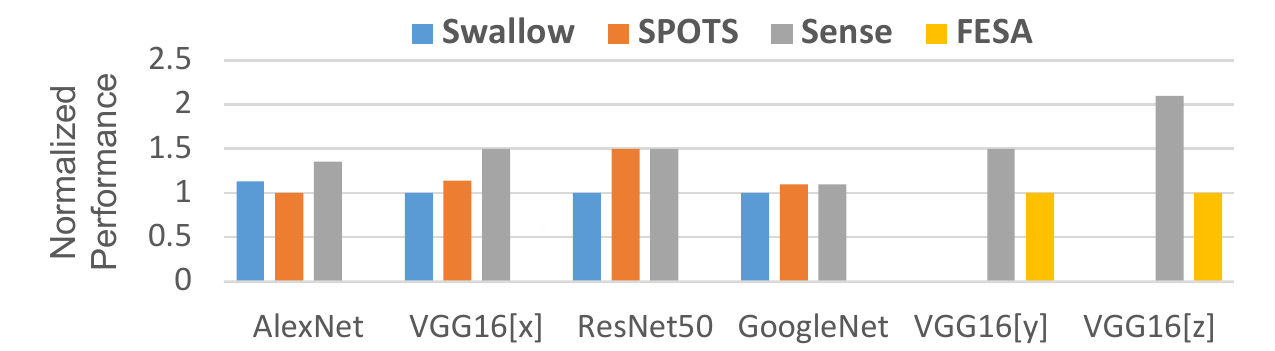}
	\caption{Performance comparison \iffalse of Sense\fi with Swallow, FESA and SPOTS in terms of sparse IFMs. \iffalse Sense achieves $1.1\times$$\textasciitilde$$1.5\times$, $1.5\times$$\textasciitilde$$2.1\times$ and $1\times$$\textasciitilde$$1.35\times$ speedup compared with Swallow, FESA and SPOTS respectively.\fi}
	\label{fig19}
\end{figure}
\begin{figure}[H]
	\centering
	\includegraphics[width=0.99\linewidth]{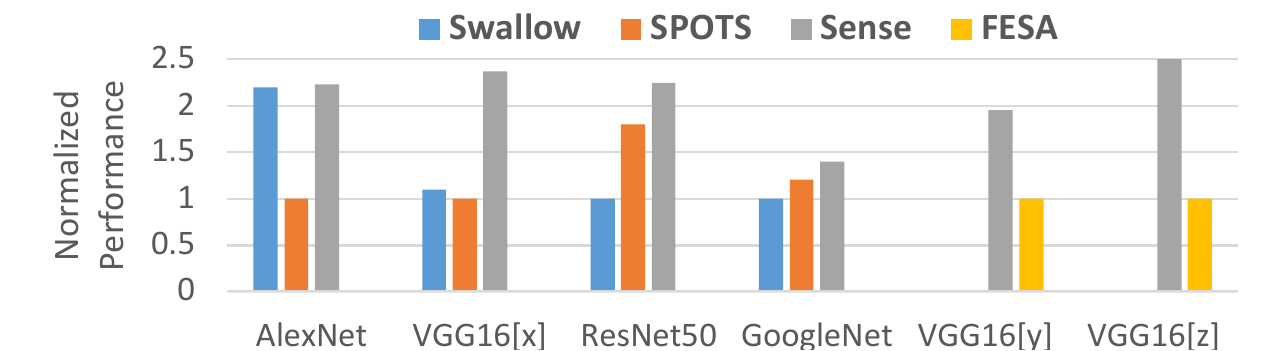}
	\caption{Overall performance comparison \iffalse of Sense\fi with Swallow, FESA and SPOTS. \iffalse Sense achieves $1\times$\textasciitilde$2.25\times$, $1.95\times$\textasciitilde$2.5\times$ and $1.17\times$\textasciitilde$2.37\times$ speedup compared with Swallow, FESA and SPOTS respectively.\fi}
	\label{fig20}
\end{figure}
On the whole, as shown in Fig.\ref{fig20}, taking sparsity of both weights and IFMs into account, we obtain $1\times$\textasciitilde$2.25\times$ and $1.17\times$\textasciitilde$2.37\times$ performance improvement on AlexNet, VGG-16, ResNet-50 and GoogleNet compared with Swallow and SPOTS respectively; and we achieve $1.95\times$\textasciitilde$2.5\times$ speedup on VGG-16 for Cifar-10 and Cifar-100 respectively compared with FESA.\\
\indent\setlength{\parindent}{1em}For PE utilization, $U_{PE}(=P_a / P_i)$, it can be calculated by the ratio of the actual performance, $P_a$ and ideal performance, $P_i(=C_c / T_p)$, where $C_c$ and $T_p$ represent computing complexity and peak throughput respectively. Since the sparsity of IFMs and weights in Sense, Swallow and FESA differs, the total computing complexity is different, making it inaccurate to analyze performance merely through PE utilization. Thus, to explore the advantages of sparse processing on PE utilization, we conduct comparison between dense systolic array and Sense based on the same data sparsity, indicating that Sense achieves $1.98\times$\textasciitilde$5.53\times$ PE utilization than dense systolic array.

\begin{figure}[H]
	\centering
	\includegraphics[width=0.9\linewidth]{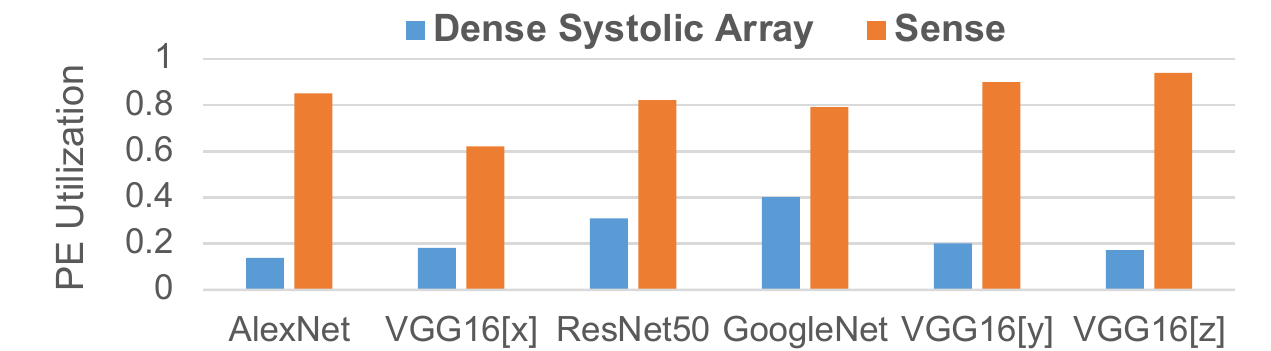}
	\caption{PE utilization comparison \iffalse of Sense\fi with dense systolic array. \iffalse Sense achieve $1.98\times$\textasciitilde$5.53\times$ PE utilization compared with dense systolic array.\fi}
	\label{fig21}
\end{figure}

\subsection{Energy Efficiency}

\begin{figure}[H]
	\centering
	\includegraphics[width=0.99\linewidth]{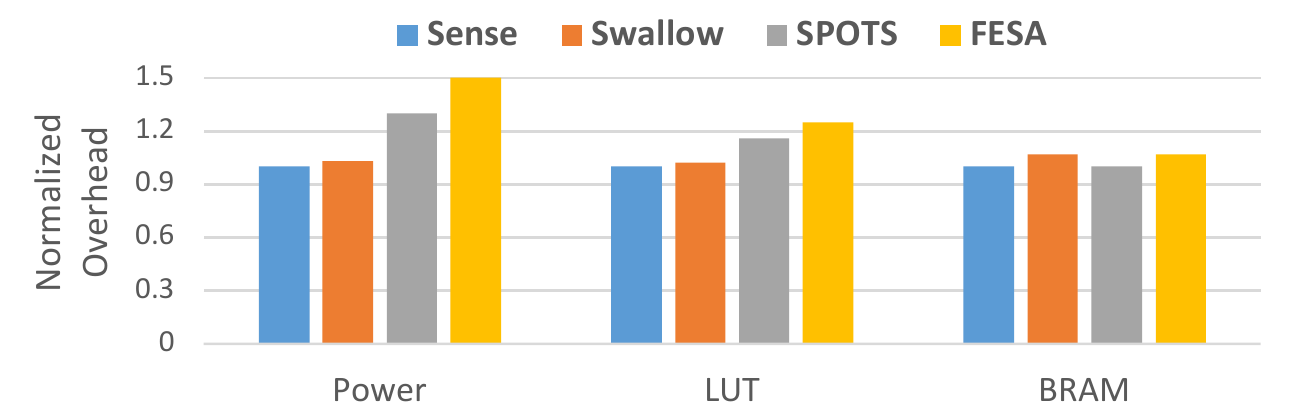}
	\caption{Power and resource overhead compared \iffalse of Sense\fi with Swallow, FESA and SPOTS. \iffalse Sense achieves $0.97\times$ and $2.25\times$, $1.3\times$ and $1.67\times$ and $0.94\times$ and $1.82\times$ energy efficiency compared with Swallow, FESA and SPOTS respectively.\fi}
	\label{fig32}
\end{figure}

\begin{figure}[H]
	\centering
	\includegraphics[width=0.99\linewidth]{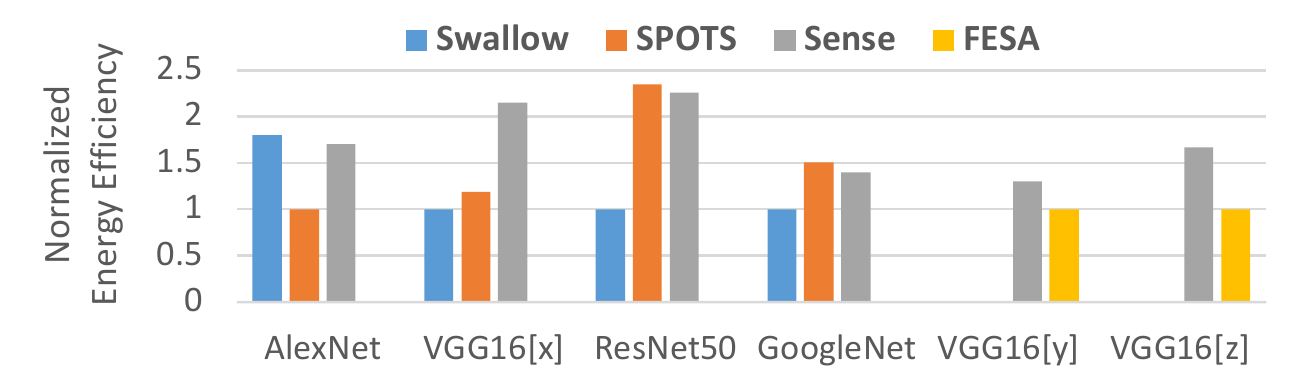}
	\caption{Energy efficiency comparison \iffalse of Sense\fi with Swallow, FESA and SPOTS. \iffalse Sense achieves $0.97\times$ and $2.25\times$, $1.3\times$ and $1.67\times$ and $0.94\times$ and $1.82\times$ energy efficiency compared with Swallow, FESA and SPOTS respectively.\fi}
	\label{fig22}
\end{figure}

Since we execute additional sparse processing compared with Swallow, FESA and SPOTS, it will incur resource and power overhead as shown in Fig.\ref{fig32}. To explore whether the benefits brought by these operations outweigh the incurred overheads, we compared the energy efficiency of Sense with Swallow, FESA and SPOTS, achieving $0.97\times$\textasciitilde$2.25\times$, $1.3\times$\textasciitilde$1.67\times$ and $0.94\times$\textasciitilde $1.82\times$ improvement respectively as shown in Fig.\ref{fig22}. For there exists difference in detail of each accelerator, it is difficult to make an absolutely fair comparison. Besides, Swallow, FESA, SPOTS and Sense are typical systolic array, which is similar in overall architecture. Thus, we only count in overhead related to sparse processing. 
Compared with Swallow, Sense further included channel clustering module, inducing $6/368 = 1.6\%$ LUT and $35/502 = 7\%$ BRAM overhead according to Tab.\ref{tab:tab5}. And this module is only active for a short period of time after OFMs finish computing, inducing 3\% extra power consumption. Thus, Sense achieves $0.97\times$\textasciitilde$2.25\times$ energy efficiency improvement compared with Swallow. \\
\indent\setlength{\parindent}{1em}For FESA, the sparse processing is only offline pruning and its dataflow is the same as dense systolic array for the regular sparsity distribution of each kernel. Thus, compared with FESA, Sense extra included Psum buffer in PE, channel clustering and compressing module, which takes up 25\% logic resource and 7\% BRAM overhead totally; since PE array is always active during execution period, Sense induced $1.5\times$ power consumption for additional sparse processing. Thus, Sense achieves $1.3\times$\textasciitilde$1.67\times$ energy efficiency improvement. \\
\indent\setlength{\parindent}{1em}As for SPOTS, the IM2COL unit exploit the temporal localities of Psum when the kernels and IFMs are transported vertically and horizontally in PE array based on the regular computing flow of GEMM, which can save the Psum buffer in each PE compared with Sense. Thus, we consume 16\% LUT resource and 30\% power more than SPOTS, obtaining $0.94\times$\textasciitilde$1.82\times$ energy efficiency.

\subsection{DRAM Access}

\begin{figure}[H]
	\centering
	\includegraphics[width=0.89\linewidth]{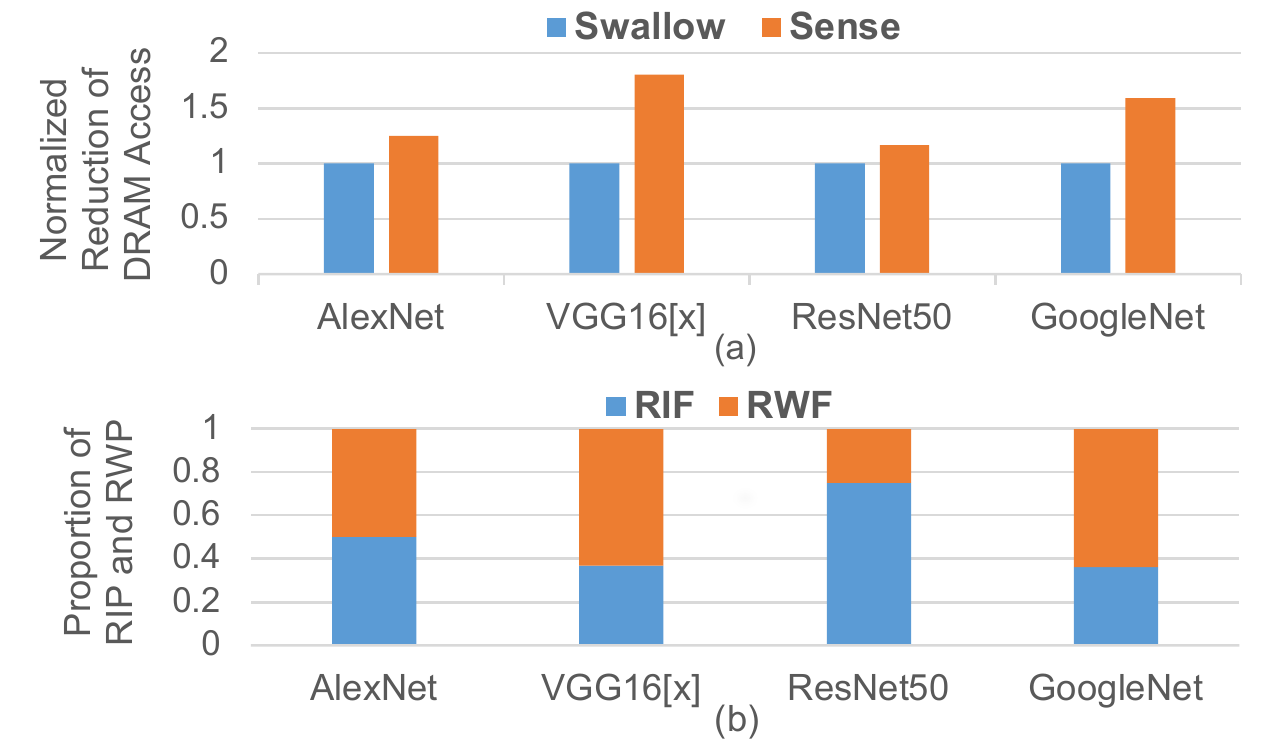}
	\caption{DRAM access comparison \iffalse of Sense \fi with Swallow. (a) DRAM access reduction.\iffalse Sense achieve $1.17\times$$\textasciitilde$$1.8\times$ reduction compared with Swallow.\fi (b) Proportion of RIF and RWF.}
	\label{fig23}
\end{figure}

The power of DRAM access takes up a large proportion in overall architecture, which can be optimized by dataflow. Since FESA and SPOTS didn't explore dataflow, we only compared DRAM access of Sense against Swallow, achieving $1.17\times$\textasciitilde$1.8\times$ reduction as shown in Fig.\ref{fig23}. For the dataflow “compute-in-row” of Swallow, it stations one row of IFMs and access weights repeatedly, which is basically "RIF" mode and inefficient for DRAM access when weights storage proportion is much larger than IFMs. Thus, Sense applied Adaptive Reusing Dataflow strategy to determine whether to "RIF or "RWF", based on their storage proportions. 
\indent\setlength{\parindent}{1em}Compared with the single dataflow of "RIF" in Swallow, we designed to  transform between "RIF" and "RWF" dataflows based on different ratios of weights and IFMs storage in each layer, further reducing DRAM access. Therefore, in VGG-16 and GoogleNet, there are nearly 60\% layers of "RWF" dataflow as shown in Fig.\ref{fig23}(b), indicating these layers can be optimized through dataflow and achieving $1.59\times$\textasciitilde$1.8\times$ DRAM access reduction. However, in AlexNet and ResNet, “RIF” already accounts for a larger proportion or there exists little difference between "RIF" and "RWF", which leads to a relatively small DRAM access reduction of $1.17\times$\textasciitilde$1.25\times$.

\subsection{Overall Comparison with Non-systolic Architectures}

\begin{figure}[H]
	\centering
	\includegraphics[width=0.99\linewidth]{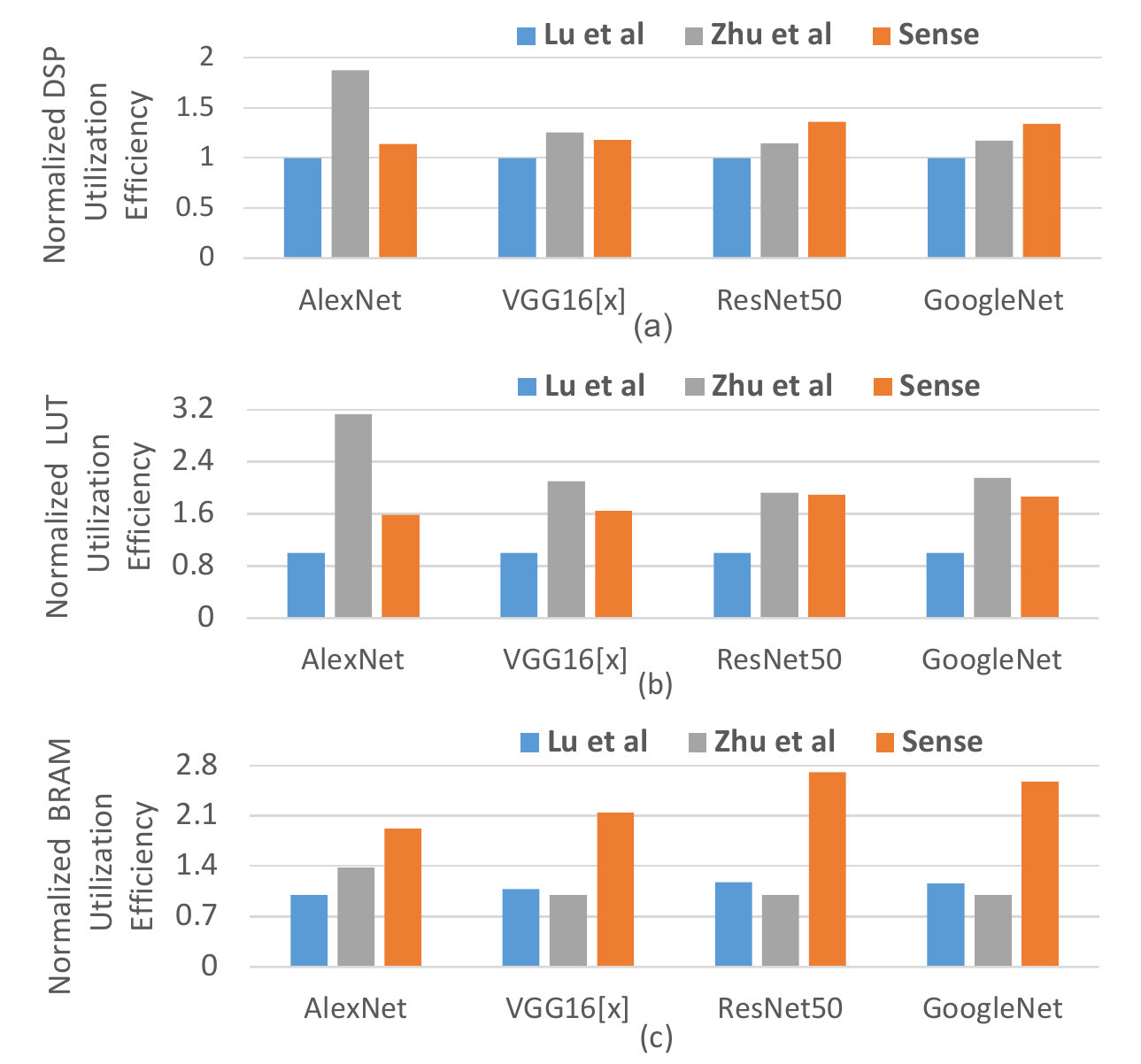}
	\caption{Resource utilization efficiency comparison \iffalse of Sense\fi with Lu $et$ $al$\cite{8735526} and Zhu $et$ $al$\cite{9130762}. (a) DSP. \iffalse comparison.   Sense achieve $1.13\times$$\textasciitilde$$1.37\times$ \iffalse and $0.7\times$$\textasciitilde$$1.17\times$ \fi DSP utilization efficiency compared with Lu $et$ $al$\cite{8735526} \iffalse and Zhu $et$ $al$\cite{9130762}\fi.\fi (b) LUT. \iffalse comparison.  Sense achieve $1.58\times$$\textasciitilde$$1.89\times$ \iffalse and $0.51\times$$\textasciitilde$$0.99\times$ \fi LUT utilization efficiency compared with Lu $et$ $al$\cite{8735526} \iffalse  and Zhu $et$ $al$\cite{9130762} \fi.\fi (c) BRAM. \iffalse comparison.  Sense achieve $1.91\times$$\textasciitilde$$2.45\times$ \iffalse and $1.36\times$$\textasciitilde$$2.7\times$ \fi BRAM utilization efficiency compared with Lu $et$ $al$\cite{8735526} \iffalse and Zhu $et$ $al$\cite{9130762}\fi.\fi}
	\label{fig24}
\end{figure}

Systolic array has intrinsic advantage of processing regular dataflow with low hardware cost, but is nowhere near customized non-systolic architectures when it comes to irregular sparse processing. However, by the treatment of load-balancing weight pruning and channel clustering, Sense achieves similar performance with customized non-systolic accelerators but consumes less resource and power. A overall comparison on Sense and previous FPGA sparse accelerators, Lu $et$ $al$\cite{8735526} and Zhu $et$ $al$\cite{9130762}, is shown in Tab.\ref{tab:tab7}. And resource utilization is analyzed in Fig.\ref{fig24}. \\
\indent\setlength{\parindent}{1em}In these two architectures, IFMs and weights are processed in each PE independently, which can naturally avoid the imbalanced loads in systolic array and achieve similar or even a little higher performance. For Lu $et$ $al$\cite{8735526}, it only addresses weight sparsity and gain no performance benefits from sparse IFMs. Therefore, Sense achieve $1.05\times$\textasciitilde$1.24\times$ speedup compared with Lu $et$ $al$\cite{8735526}. For Zhu $et$ $al$\cite{9130762}, it exploit both IFM and weihgts for acceleration. But its PE number is $1.3\times$ more than Sense, therefore achieving $1.3\times$ peak throughput. Thus, Zhu $et$ $al$\cite{9130762} obtained $1.06\times$\textasciitilde$1.35\times$ performance on VGG-16, ResNet-50. For AlexNet, Zhu $et$ $al$\cite{9130762} is $2.1\times$ faster than Sense because its sparsity of IFMs is 35\% higher than us, which cannot be changed by software or hardware. \\
\begin{table*}\centering
	\caption{Overall Comparison \iffalse of Sense \fi With Lu $et$ $al$\cite{8735526} and Zhu $et$ $al$\cite{9130762}}
	\label{tab:tab7}
	\begin{tabular}{|c|c|c|c|cccc|}
		\hline
		\multirow{2}{*}{Accelerator} & \multirow{2}{*}{DSP/LUT/BRAM} & \multirow{2}{*}{Power} & \multirow{2}{*}{\begin{tabular}[c]{@{}c@{}}Peak Throughput\\ (GMAC/s)\end{tabular}} & \multicolumn{4}{c|}{Performance(image/s)/Energy Efficiency(image/J)}                                      \\ \cline{5-8} 
		&                      &                        &                                                                                & \multicolumn{1}{c|}{AlexNet}  & \multicolumn{1}{c|}{VGG-16}  & \multicolumn{1}{c|}{ResNet-50} & GoogleNet \\ \hline
		Lu $et$ $al$\cite{8735526}                 & 1144/552K/912        & 23.6W                  & 204.8                                                                          & \multicolumn{1}{c|}{446/18.9} & \multicolumn{1}{c|}{31/1.3}  & \multicolumn{1}{c|}{42/1.8}    & 154/6.5   \\ \hline
		Zhu $et$ $al$\cite{9130762}                & 1352/390K/1460       & 15.4W                  & 268.8                                                                          & \multicolumn{1}{c|}{987/64.1} & \multicolumn{1}{c|}{46/2.99} & \multicolumn{1}{c|}{57/3.7}    & 215/14.0  \\ \hline
		Sense                        & 1061/348K/502        & 10.8W                  & 204.8                                                                          & \multicolumn{1}{c|}{471/43.6} & \multicolumn{1}{c|}{34/3.15} & \multicolumn{1}{c|}{53/4.9}    & 191/17.7  \\ \hline
	\end{tabular}
\end{table*}
\indent\setlength{\parindent}{1em}However, independent data supply of each PE costs huge overhead. In Lu et al\cite{8735526}, unlike simple systolic dataflow, a large amount of additional logic and BRAM resources are consumed by CMUX and TLUT modules because of the complex connection between PE and output buffer for index matching. And Zhu et al\cite{9130762} requires large BRAM resources for OFM storage in each PE, while Sense calculates OFMs in each column. The resource utilization efficiency(performance/logic cell) is shown in Fig.\ref{fig24}. Compared with Lu $et$ $al$\cite{8735526}, Sense is $1.13\times$\textasciitilde$1.37\times$, $1.66\times$\textasciitilde$1.98\times$ and $1.91\times$\textasciitilde$2.45\times$ more efficient on DSP, LUT and BRAM utilization respectively. Compared with Zhu $et$ $al$\cite{9130762}, expect for AlexNet, Sense achieves $0.94\times$\textasciitilde$1.17\times$, $0.82\times$\textasciitilde$1.04\times$ and $2.15\times$\textasciitilde$2.7\times$ resources utilization efficiency on DSP, LUT and BRAM respectively, indicating that we obtain approximate resource utilization on LUT and DSP, and outstand on BRAM. As for AlexNet, the performance benefits of Zhu $et$ $al$\cite{9130762} far outweigh our hardware saving, resulting in $1.6\times$ and $2\times$ than Sense on DSP and LUT utilization efficiency.

With more resource comes more power. Consequently, Sense consumes $2.15\times$ and $1.43\times$ less power than Lu $et$ $al$\cite{8735526} and Zhu $et$ $al$\cite{9130762} respectively. Though the power of Sense is simulated by Vivado, it still can be a reference to power comparison. Taking the analysis of power, resource and performance all into account, Sense gains $2.31\times$\textasciitilde$2.75\times$, $0.71\times$\textasciitilde$1.37\times$ energy efficiency improvement compared with Lu $et$ $al$\cite{8735526} and Zhu $et$ $al$\cite{9130762} respectively as shown in Tab.\ref{tab:tab7}. Besides, when it comes to dense processing, our power consumption is $2.79\times$ and $1.86\times$  less than Lu $et$ $al$\cite{8735526} and Zhu $et$ $al$\cite{9130762} respectively, achieving much higher energy efficiency.

\subsection{Design Space Exploration}

\begin{figure}[H]
	\centering
	\includegraphics[width=0.99\linewidth]{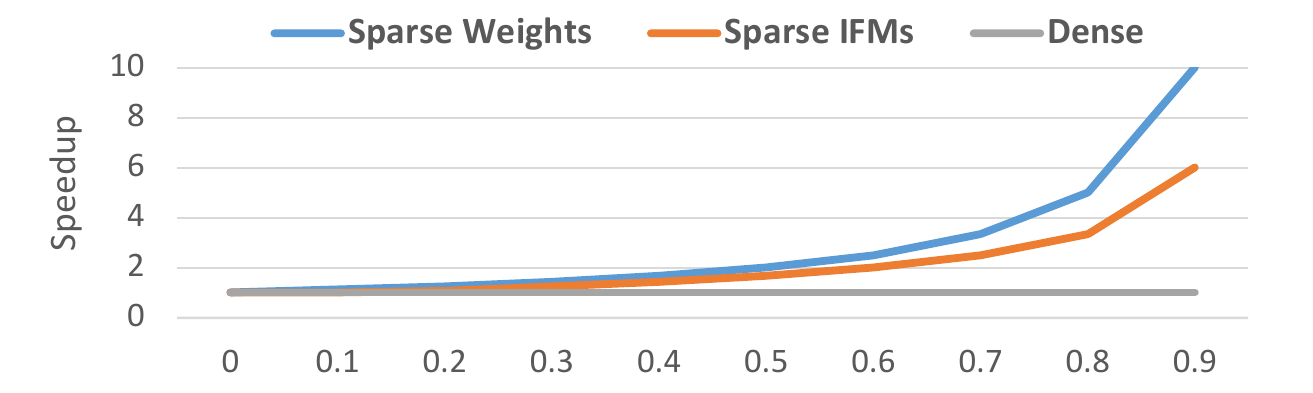}
	\caption{Speedup by exploiting IFM and weights of different sparsity separately.}
	\label{fig26}
\end{figure}

\begin{figure}[H]
	\centering
	\includegraphics[width=0.99\linewidth]{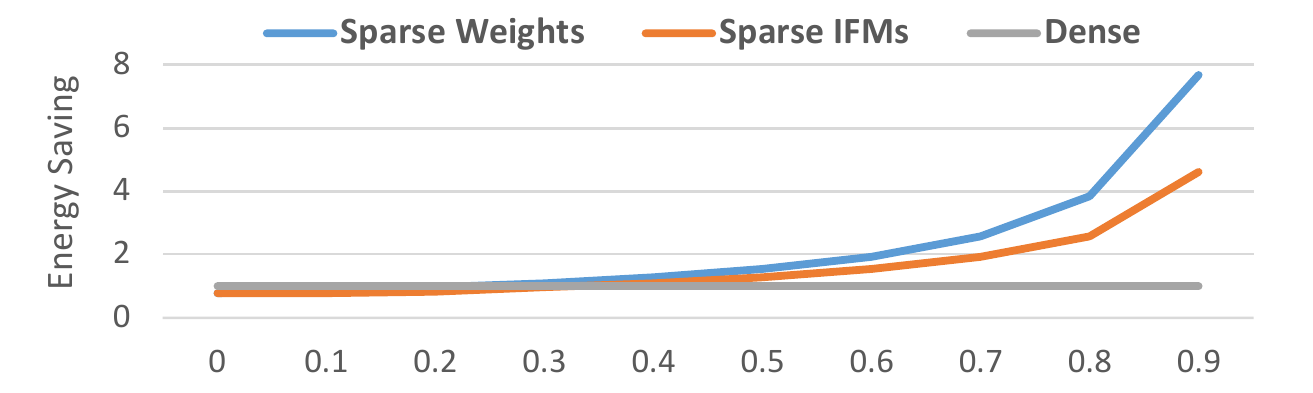}
	\caption{Energy saving by exploiting IFM and weights of different sparsity separately.}
	\label{fig27}
\end{figure}

1)Sensitivity to Sparsity. Sparse processing greatly accelerates computation, but comes with extra operations, inducing 30\% power overhead. Based on the influence of sparsity  ratio on performance and power, we determine a threshold for computing mode converting, whether dense or sparse. In this way, our architecture can handle data of any ratio of sparsity with higher energy efficiency. \\
\indent\setlength{\parindent}{1em}Taking sparsity ratio of 10\% as stride, we achieve different speedup and energy saving by exploiting IFMs and weights of different sparsity ratios separately, as shown in Fig.\ref{fig26} and Fig.\ref{fig27} respectively. Weight sparsity can be 100\% exploited for acceleration based on load-balancing weight pruning, achieving $1\times$\textasciitilde$ 10\times$ speedup. Since IFM sparsity cannot be constrained by offline training, the efficiency of sparsity exploiting on IFMs is a bit lower than weights, achieving $1\times$\textasciitilde$ 6\times$ speedup. For energy saving, it can be calculated by the multiplication of normalized power and speedup, indicating $0.77\times$\textasciitilde$ 7.69\times$ energy reduction on weights and $0.77\times$\textasciitilde$ 4.62\times$ on IFMs. 

\begin{figure}[H]
	\centering
	\includegraphics[width=0.99\linewidth]{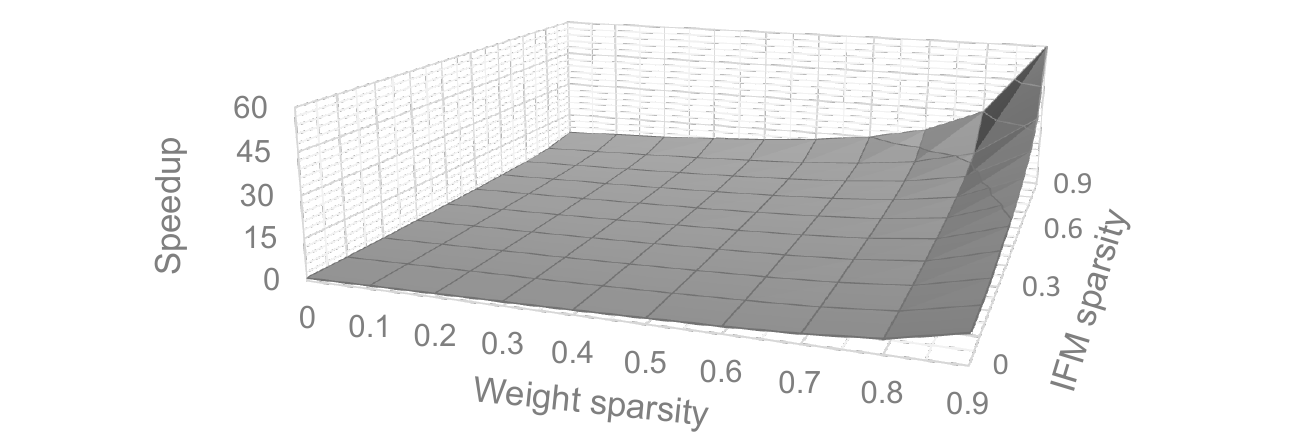}
	\caption{Speed up by exploiting both IFM and weights of different sparsity.}
	\label{fig28}
\end{figure}

\begin{figure}[H]
	\centering
	\includegraphics[width=0.99\linewidth]{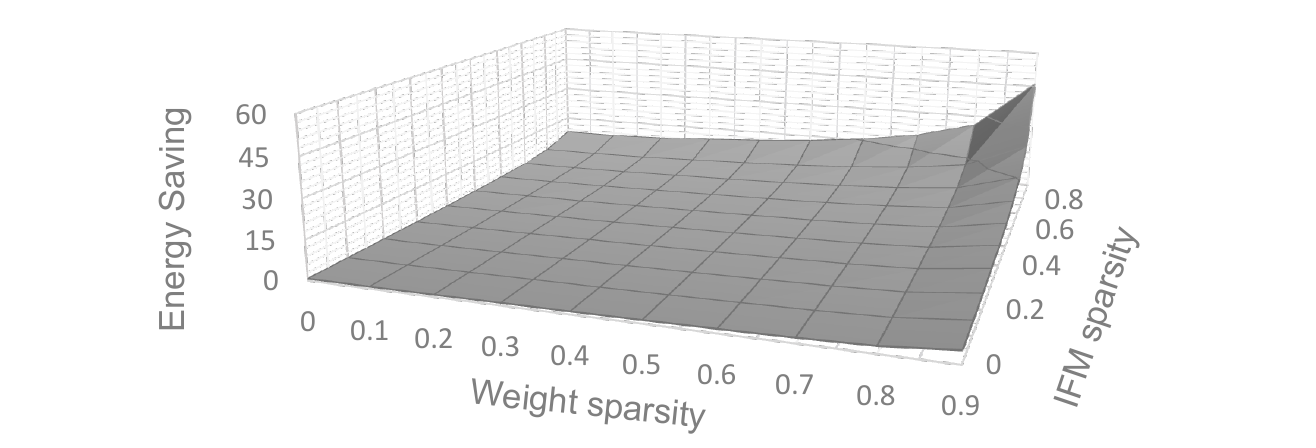}
	\caption{Energy saving by exploiting both IFM and weights of different sparsity.}
	\label{fig29}
\end{figure}

Fig.\ref{fig28} and Fig.\ref{fig29} show the speedup and energy saving by exploiting different sparsity of both IFM and weights. When sparsity ratios of IFM and weight are beyond 30\% and 20\% respectively, sparse computing mode is more energy efficient. Therefore, we combined dense and sparse processing and computing modes are switched based on sparsity ratios.

\begin{figure}[H]
	\centering
	\includegraphics[width=0.99\linewidth]{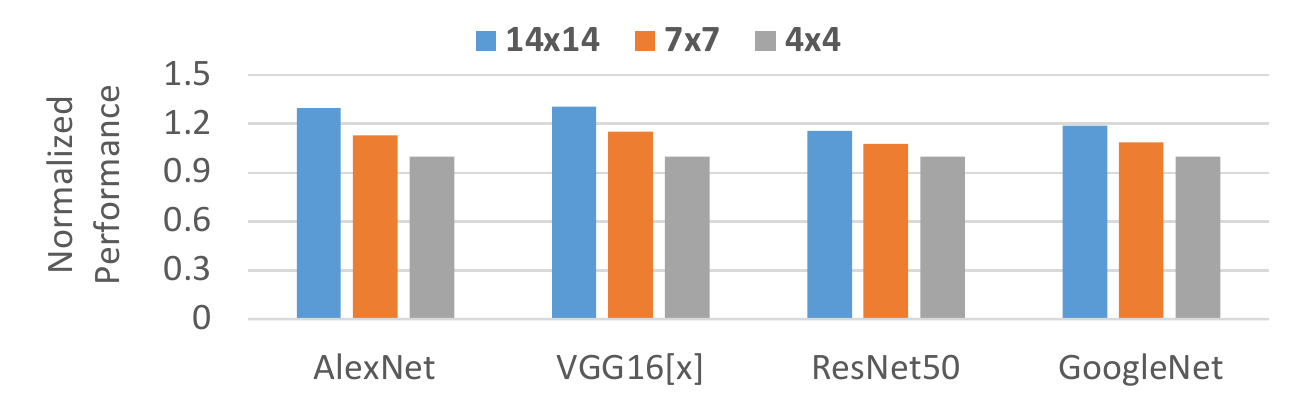}
	\caption{Normalized Performance of channel clustering based on different size of PE array.}
	\label{fig30}
\end{figure}

2)Hardware effect on efficiency of channel clustering. Since channel clustering is performed on hardware, the efficiency of performance improvement mainly depends on the size of PE array , $N_{PE}$ , and size of IFM sub-tiles, $N_{is}$. For size of PE array, we process $N_{PE}$ IFM simultaneously, so the computing time is determined by the IFMs with lowest sparsity based on systolic dataflow. The larger $N_{PE}$ is, the more likely the PE array contains IFM with relatively low sparsity, which can block performance improvement. Thus, we set three different $N_{PE}$, $8\times8$, $16\times16$ and $32\times32$, to explore their performance on different CNNs. As shown in Fig.\ref{fig30}, the performance of $8\times8$ is $1.05\times$\textasciitilde$ 1.07\times$ and $1.1\times$\textasciitilde$ 1.15\times$ higher than $16\times16$ and $32\times32$ respectively, indicating that PE array size have little effect on performance. The main reason for this phenomenon is that PE array of small size only outstands on layers with small IC numbers, while the size of IC becomes bigger as the layer goes deeper in these CNNs. Thus, we choose $32\times32$ PE array for higher resource utilization of FPGA with little performance sacrifice.  

\begin{figure}[H]
	\centering
	\includegraphics[width=0.99\linewidth]{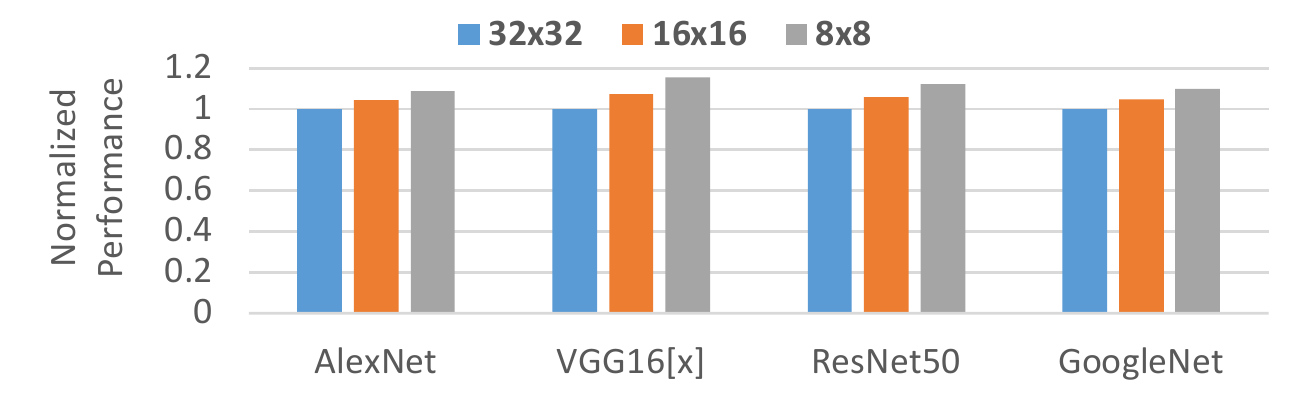}
	\caption{Normalized Performance of channel clustering based on different size of IFM sub-tiles.}
	\label{fig31}
\end{figure}

For size of IFM sub-tiles, since we gather IFMs with approximate sparsity by sorting the number of NZEs in each IFM, the distribution of sparsity of each sub-tile in the IFM can be still irregular, causing imbalanced work load when there still exists sparsity difference in gathered IFM sub-tiles. The smaller the size of IFM sub-tiles is, the greater chance of imbalance workload there can be. Thus, to explore the effect of $N_{is}$ on performance, we set three different $N_{is}$, $14\times14$, $7\times7$ and $4\times4$ and verify on different CNNs. As shown in Fig.\ref{fig31}, the performance of $14\times14$ is $1.1\times$\textasciitilde$ 1.18\times$ and $1.3\times$\textasciitilde$ 1.32\times$ higher than $7\times7$ and $4\times4$ on AlexNet and VGG-16 respectively. For ResNet and GoogleNet, the improvement reduces to nearly $1.1\times$ because the IFM size in each layer are relatively small. Therefore, we we choose $7\times7$ sub-tiles for the balance of performance and hardware overhead.

\section{Conclusion}
This paper proposed a sparse systolic-array-based accelerator, called Sense, for both sparse IFM and weight processing, achieving large performance improvement with relatively small resource and power consumption. Meanwhile, we applied channel clustering  to gather IFMs with approximate sparsity and co-designed a load-balancing weight pruning method to keep the sparsity ratio of each kernel at a certain value with little accuracy loss. This treatment can effectively balance workload of sparse IFMs and weights in systolic array. Additionally, Adaptive Dataflow Configuration is applied to map various dataflows on Sense, enhancing data reuse, lowering DRAM access and further reducing system energy consumption. Compared with sparse systolic accelerator, Sense can achieve higher performance and energy efficiency with reasonable overhead and meanwhile maintain versatility; compared with non-systolic sparse accelerators, Sense can achieve higher resource utilization efficiency.

\bibliographystyle{IEEEtran}
\bibliography{TVLSI}

\end{document}